\documentclass[a4paper,12pt]{article}
\pdfoutput=1
\pdfoutput=1
\usepackage{jcappub}
\usepackage{amsthm}
\usepackage{graphicx}
\usepackage{color}
\usepackage{amsmath}
\usepackage{graphicx,textcomp,float,gensymb,wrapfig, enumitem,comment,dsfont,subfigure,framed,slashed,appendix,wrapfig,wasysym}
\usepackage{comment}

\newcommand{\be}{\begin{equation}}
\newcommand{\ee}{\end{equation}}
\def\bsp#1\esp{\begin{split}#1\end{split}}

\newcommand{\bea}{\begin{eqnarray}}  
\newcommand{\eea}{\end{eqnarray}}  

\title{Probing the sterile neutrino portal to Dark Matter with $\gamma$ rays}
\author[a]{Miguel G. Folgado,}
\author[b]{Germ\'an A. G\'omez-Vargas,}
\author[a]{Nuria Rius}
\author[a]{and Roberto Ruiz de Austri}
\affiliation[a]{Departamento de F\'isica Te\'orica and IFIC, Universidad de Valencia-CSIC,
C/ Catedr\'atico Jos\'e Beltr\'an, 2, E-46980 Paterna, Spain}
\affiliation[b]{Instituto de Astrof\'\i sica, Pontificia Universidad Cat\'olica de Chile, Avda. Vicu\~na
Mackenna 4860, Santiago, Chile}

\emailAdd{migarfol@ific.uv.es}
\emailAdd{gggomezv@uc.cl}
\emailAdd{nuria.rius@ific.uv.es}
\emailAdd{rruiz@ific.uv.es}


\abstract{Sterile neutrinos could provide a link between the Standard Model particles and a dark sector, besides generating active neutrino masses via the seesaw mechanism type I. We show that, if dark matter annihilation into sterile neutrinos determines its observed relic abundance, it is possible to explain the Galactic Center $\gamma$-ray excess reported by the Fermi-LAT Collaboration as due to an astrophysical component plus dark matter annihilations. We observe that sterile neutrino portal to dark matter provides an impressively good fit, with a p-value of 0.78 in the best fit point, 
to the Galactic Center $\gamma$-ray flux, for DM masses in the range (40-80) GeV and sterile neutrino masses 20 GeV $\lesssim M_N < M_{DM}$. 
Such values are compatible with the limits from Fermi-LAT observations of the dwarfs spheroidal galaxies in the Milky Way halo, which rule out dark matter masses below $\sim$ 50 GeV (90 GeV), for sterile neutrino masses $M_N \lesssim M_{DM}$ ($M_N \ll M_{DM}$). We also estimate the impact of AMS-02 anti-proton data on this scenario.}

\begin{document}
\hfill {\tt FTUV-18-0226.9011, IFIC/18-10}

\maketitle
\flushbottom
 
\section{Introduction}
Dark matter (DM) and neutrino masses constitute indubitable observational evidence for physics beyond the Standard Model (SM) of fundamental interactions. Thus, the existence of a connection between the new degrees of freedom needed to account for both observations is an exciting possibility to explore. In particular, if DM is a thermal relic of the early Universe and the {\it seesaw mechanism} is realized to generate neutrino masses, new massive particles are required to solve both problems. 
The most economical scenario, namely that the sterile neutrinos constitute the DM \cite{Dodelson:1993je}, has been thoroughly studied  \cite{Adhikari:2016bei}. Hence we consider in this work a different case: The sterile neutrino portal to DM. In this scenario DM is an SM singlet state that interacts mainly with sterile neutrinos, being such interactions of the right strength to produce the observed DM relic abundance \cite{Pospelov:2007mp,Escudero:2016tzx,Escudero:2016ksa}.  

DM interactions with SM particles are very weak to avoid collider and direct detection constraints, although they must reproduce the correct abundance of DM thermally through its annihilation into sterile neutrinos which eventually decay into SM particles. This decay is due to Yukawa couplings of sterile neutrino and leptons which also generate a Majorana mass for the light neutrinos via the type I seesaw mechanism. In general, if DM s-wave interactions dominate the annihilation process, we expect to have indirect detection signals, searches for these signals lead to the most stringent bounds on this scenario \cite{Escudero:2016ksa}. 

A comprehensive analysis of indirect detection hunts within the sterile neutrino portal to DM has been presented in \cite{Batell:2017rol}, including constraints from Planck CMB measurements, $\gamma$-ray flux collected by the Fermi Large Area Telescope (LAT), and AMS-02 antiproton observations. 
Indirect signals from solar DM annihilation to long-lived sterile neutrinos have been analyzed in \cite{Allahverdi:2016fvl}.
The primary target for neutral DM annihilation products is the Galactic Center, as we expect there the largest DM concentration in the nearby cosmos. Interestingly, an unexpected signal detected in the gamma-ray data collected by the Fermi LAT from the inner Galaxy, the so-called Galactic Center Excess (GCE). It has created a great excitement because its spectral energy distribution and morphology are consistent with predictions from DM annihilation\cite{Goodenough:2009gk, Vitale:2009hr, Hooper:2010mq, Gordon:2013vta, Hooper:2011ti, Daylan:2014rsa, 2011PhLB..705..165B, Calore:2014xka, Abazajian:2014fta,Zhou:2014lva}. All those works devoted to analyzing the GCE confirm that its properties strongly depends on the analysis method used to subtract it from the Fermi-LAT data. The variation in the GCE properties with the analysis causes modifications in the models able to explain it. The work in \cite{Tang:2015coo} shows that it is possible to account for the GCE obtained in \cite{Calore:2014xka} by DM annihilation into sterile neutrinos. In \cite{Batell:2017rol} the compatibility of the GCE DM interpretation with the other indirect searches is discussed. The dwarf spheroidal galaxies (dSphs) are pristine targets for DM signals because they lack detectable gamma-ray sources. The authors of \cite{Campos:2017odj} use the Fermi-LAT gamma-ray data from dSphs to set limits on DM annihilations into sterile neutrinos.

In this paper we consider a new Fermi-LAT analysis of Pass 8 data on Galactic Center $\gamma$-rays presented in \cite{TheFermi-LAT:2017vmf}, and we explore the ability of the DM sterile neutrino portal to account for the GCE, which is peaked at $\sim$ 3 GeV, that is, slightly higher energies than reported in previous analysis. We also derive the limits from dSphs. 
Although we use a particular realization of the sterile neutrino portal DM, the results of our analysis can be applied to other models, provided the sterile neutrino decays only to SM particles.

The paper is organized as follows. In Sec.~\ref{sec:snp} we briefly review the sterile neutrino portal scenario, and derive the SM particle spectra from sterile neutrino decays,
relevant for the indirect detection constraints on such portal. In Sec.~\ref{sec:gc} we describe the model independent fit to the GCE, while in Sec.~\ref{sec:dSphs} we present the limits from Fermi-LAT dSphs and AMS-02 anti-proton data. We conclude in Sec.~\ref{sec:con}. 

\section{Sterile neutrino portal to Dark Matter}\label{sec:snp} 
Our analysis can be applied to any type of sterile neutrino portal scenario up to the following requirement: The observed DM relic abundance is determined by its interactions with sterile neutrinos, which in turn generate light neutrino masses via the type I seesaw mechanism. For definiteness 
in this section we consider a very simple realization studied in \cite{Escudero:2016ksa}. Besides the sterile neutrinos, the SM is extended by a dark sector that contains a scalar field 
$\phi$ and a fermion $\Psi$. These fields are both singlets of the SM gauge group but charged under a dark sector symmetry group, $ G_{dark}$, such that the combination $ \overline{\Psi} \phi$ is a singlet of this hidden symmetry. 

The lighter of the two dark particles ($\phi$ and $\Psi$) turns out to be stable if all SM particles, as well as the sterile neutrinos, are singlets of $ G_{dark}$, disregarding the nature of the dark group. The stable particle is a good DM candidate. We assume for simplicity that the dark symmetry $G_{dark}$ is a global symmetry at low energies, although our analysis is equally valid whether it is local.
  
The relevant terms of the Lagrangian are:
\bea
\mathcal{L} &=& \mu_H^2 H^\dagger H - \lambda_H (H^\dagger H)^2  -
\mu_\phi^2 \phi^\dagger \phi - \lambda_\phi (\phi^\dagger \phi)^2 - 
\lambda_{H\phi} (H^\dagger H) \, (\phi^\dagger \phi)   
\nonumber \\
&-&\left( \phi \overline{\Psi}(  \lambda_a +  \lambda_p \gamma_5) N 
 + Y \overline {L}_L H N_{R}  + {\rm h.c.}  \right)  \, 
\eea
where we have omitted flavour indexes.
The Yukawa couplings $Y$ between the right-handed fermions $N_R$ and the SM leptons lead to masses for the active neutrinos after electroweak symmetry breaking, via type I seesaw mechanism. Although at least two sterile neutrinos are required to generate the neutrino masses observed in oscillations, in our analysis we consider that only one species is lighter than the DM and therefore relevant for the determination of its relic abundance and indirect searches. The results can be easily extended to the case of two or more sterile neutrinos lighter than the DM.

Assuming that the dark fermion $\Psi$ is Majorana and 
constitutes the DM, its annihilation cross section into sterile neutrinos is given by
\footnote{Were $\Psi$ a Dirac fermion,  the exchange
$\alpha \leftrightarrow \beta$ should be performed in eq.(\ref{eq:acs}).}
\bea
\sigma v =\frac{ (\alpha + \beta \, r_{N \Psi})^2}{4 \pi M_\Psi^2 }   \, \frac{ \sqrt{1-r_{N \Psi}^2}}{(1+r_\phi^2-r_{N \Psi}^2)^2} + {\cal O}(v^2) \,
\label{eq:acs}
\eea
where $\alpha= \lambda_s^2-\lambda _p^2 $ and $\beta=\lambda_s^2+\lambda _p^2$, $r_{\phi} = M_\phi/M_{\Psi}$, and $r_{N \Psi} = M_N/M_\Psi$ and $v$ is the relative velocity of the DM particles. 
In the following, we restrict ourselves 
to a scalar interaction between the dark fermion and the $N's$, but from eq.~(\ref{eq:acs}) it is clear that a pseudoscalar coupling $\lambda_p \gamma_5$ leads to the same results. Only a chiral interaction gives rise to 
reduced indirect detection signals, since for $M_N \ll M_\Psi$ the 
annihilation cross section is effectively  
p-wave, and therefore velocity suppressed.

In the scenario presented above it is always possible to obtain the observed DM relic abundance when  $M_N < M_\Psi$ in the range $M_\Psi \in$ [1 GeV, 2 TeV] with perturbative couplings $\lambda_s \equiv \lambda \sim$ 0.01 - 1 and mediator masses $M_\phi \in$ [1 GeV, 10 TeV] \cite{Escudero:2016ksa}. 
It is worth noticing that for sufficiently small Yukawa couplings of the sterile neutrinos, it could happen that the DM $\Psi$ and $N$ bath decouple from the SM after the decay of the dark scalar, $T \lesssim M_\phi$, and remain in thermal equilibrium but with a different temperature. In this case, the DM freeze-out leads to a larger relic abundance, so that a larger annihilation cross section (and thus a larger coupling between DM and sterile neutrinos) is needed to reproduce the observed value \cite{Tang:2016sib, Bernal:2017zvx}. In Sec.~\ref{sec:dSphs} we will see that the Fermi-LAT data from dSphs can set stringent constraints on these scenarios.

If the scalar $\phi$ were the DM instead, the corresponding annihilation
cross section is very similar to eq.(\ref{eq:acs}), including the fact that it becomes velocity suppressed for $M_N \ll M_\phi$  if the DM couplings are chiral. In
\cite{Escudero:2016ksa} it has been shown that for scalar DM it is also possible to get the correct relic abundance in a comparable region of the parameter space, therefore our analysis applies to such scenario as well.

The indirect detection signatures depend on the thermally averaged total annihilation cross section, $\langle\sigma v\rangle$ (for a detailed calculation of the thermal average see for instance ref.\cite{Gondolo:1990dk}), and on the energy spectrum of the final SM particles, which is determined by $M_\Psi$ and $M_N$. Moreover, given a pair of values ($M_\Psi, M_N$), it is always possible to obtain a certain value of the cross section by appropriately choosing the other two free variables, $\lambda, M_\phi$, with the only limitation of the coupling $\lambda$ to remain perturbative. Therefore, in the next sections we will consider as free parameters ($\langle\sigma v\rangle, M_\Psi, M_N$); in this way, our analysis is valid for any other neutrino portal scenario able to reproduce the same annihilation cross section, provided the sterile neutrinos decay only to SM particles.

Light neutrino masses are generated via TeV scale type I seesaw mechanism. We denote $\nu_\alpha$ the active neutrinos and $N_s$ the sterile ones. After electroweak symmetry breaking, the neutrino mass matrix in the basis $(\nu_\alpha, N_s)$ is given by
\be
\label{eq:numass}
{\cal M}_\nu = \left(
\begin{array}{cc}
0 & M_D \\
M_D^T & M_N
\end{array}
\right) \, 
\ee
where $M_D = Y v_H/\sqrt{2}$ and $Y_{\alpha s}$ are the Yukawa couplings. The matrix ${\cal M}_\nu$ can be diagonalized by a unitary matrix $U$, so that 
\be
{\cal M}_\nu = U^* \, Diag(M_\nu,M) \, U^\dagger \, 
\ee
where $M_\nu$ is the diagonal matrix with the three lightest eigenvalues of ${\cal M}_\nu$,  
of order $M_D^2/M_N$, and $M$ contains the heavier ones, of order $M_N$. 
 
The mass eigenstates ${\bf n}=(\nu_i,N_h$) are related to the active and sterile neutrinos, 
($\nu_\alpha$, $N_{s}$), by
\bea
\left(\begin{array}{c}\nu_\alpha \\ N_s \end{array}\right)_L  =  U^* \,  
\left(\begin{array}{c}\nu_i  \\ N_h \end{array}\right)_L  \ .
\eea

The unitary matrix $U$ can be written as 
\be 
\label{eq:mixing}
U = \left(
\begin{array}{cc}
U_{\alpha i } & U_{\alpha h}  \\
U_{s i } & U_{s h } 
\end{array}
\right) \, 
\ee
where, at leading order in the seesaw expansion parameter, ${\cal O}(M_D/M_N)$:
\bea
U_{\alpha i } &=& [U_{PMNS} ]_{\alpha i} \qquad   U_{sh} = I 
\nonumber \\
U_{\alpha h } &=&  [M_D M_N^{-1}]^*_{\alpha h}
\\
U_{s i} &= & - [M_N^{-1} M_D^T \, U_{PMNS}]_{si} \ .
\nonumber 
\eea
Notice that at this order the states $N_h$ and $N_s$ coincide, therefore we identify them in the rest of this paper. 

Sterile neutrinos are produced in DM annihilations and then decay into SM particles. The decay channels depend on the sterile neutrino mass. Namely if the right-handed neutrino is lighter than the $W$ boson, $N$ will decay through off-shell $h,Z,W$ bosons to three fermions. Since the decay via a virtual $h$ is further suppressed by the small Yukawa couplings of the SM fermions, it is a very good approximation to consider only the processes mediated by virtual $W,Z$, whose partial widths read~\cite{GonzalezGarcia:1990fb}:
\bea 
 \Gamma(N  \rightarrow \nu q \bar{q} ) &=& 3\, A C_{NN}
[2(a_u^2 + b_u^2) + 3(a_d^2 + b_d^2)] f(z) 
\\
 \Gamma(N   \rightarrow  3 \nu) &=& A C_{NN}
[\frac 3 4 f(z)  +  \frac 1 4  g(z,z)]  
\\
 \Gamma(N   \rightarrow \ell q \bar{q} ) &=& 6\, A C_{NN}
 f(w,0) 
\\
 \Gamma(N   \rightarrow \nu \ell\bar{\ell} ) &=&  A C_{NN}
[3(a_e^2 + b_e^2) f(z)  + 3 f(w) - 2 a_e g(z,w) ] \, 
\eea
where 
\be
A \equiv  \frac {G_F^2 M_{N}^5 }{192 \, \pi^3}  \ , C_{ij} = \sum_{\alpha=1}^{3} U_{\alpha i} U^*_{\alpha j} \, 
\ee
$a_f,b_f$ are the left and right neutral current couplings of the fermions ($f=q,\ell$),  the variables $z,w$ are given by 
\be 
z= (M_N/M_Z)^2  \ , \qquad w = (M_N/M_W)^2 \, 
\ee 
and the functions $f(z), f(w,0)$ and $g(z,w)$ can be found in \cite{Dittmar:1989yg}.

On the other hand, if $M_N > M_W$ two body decays to SM particles are open, and the corresponding widths are \cite{Pilaftsis:1991ug}:
\bea 
  \label{eq:WZh}
  \Gamma(N  \rightarrow W^{\pm} \ell^{\mp}_{\alpha}  ) &=& \frac{g^2}{64 \pi} |U_{\alpha N}|^2 
  \frac{M_N^3}{M_W^2} \left(1 -   \frac{M_W^2}{M_N^2} \right)^2
   \left(1 +  \frac{2 M_W^2}{M_N^2} \right)  
   \\
    \Gamma(N  \rightarrow Z  \, \nu_\alpha ) &=& \frac{g^2}{64 \pi c_W^2 } |C_{\alpha N}|^2 
  \frac{M_N^3}{M_Z^2} \left(1 -   \frac{M_Z^2}{M_N^2} \right)^2
   \left(1 +  \frac{2 M_Z^2}{M_N^2} \right)    
   \\
   \Gamma(N  \rightarrow h \,  \nu_\alpha ) &=& \frac{g^2}{64 \pi } |C_{\alpha N}|^2 
  \frac{M_N^3}{M_W^2} \left(1 -   \frac{M_h^2}{M_N^2} \right)^2 \ .
\eea

\begin{figure}[htbp]
\centering
\subfigure{\includegraphics[width=76mm]{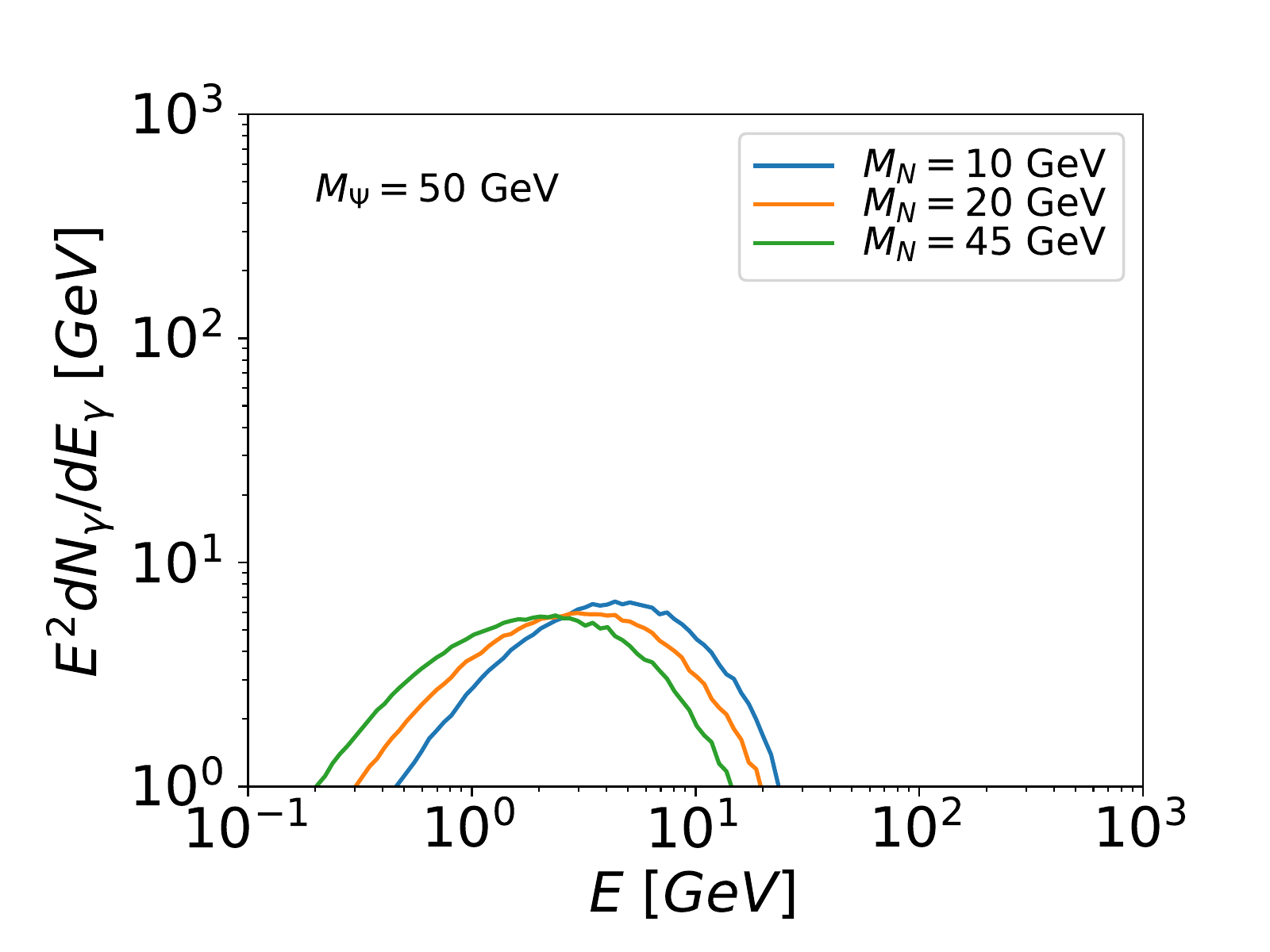}}
\subfigure{\includegraphics[width=76mm]{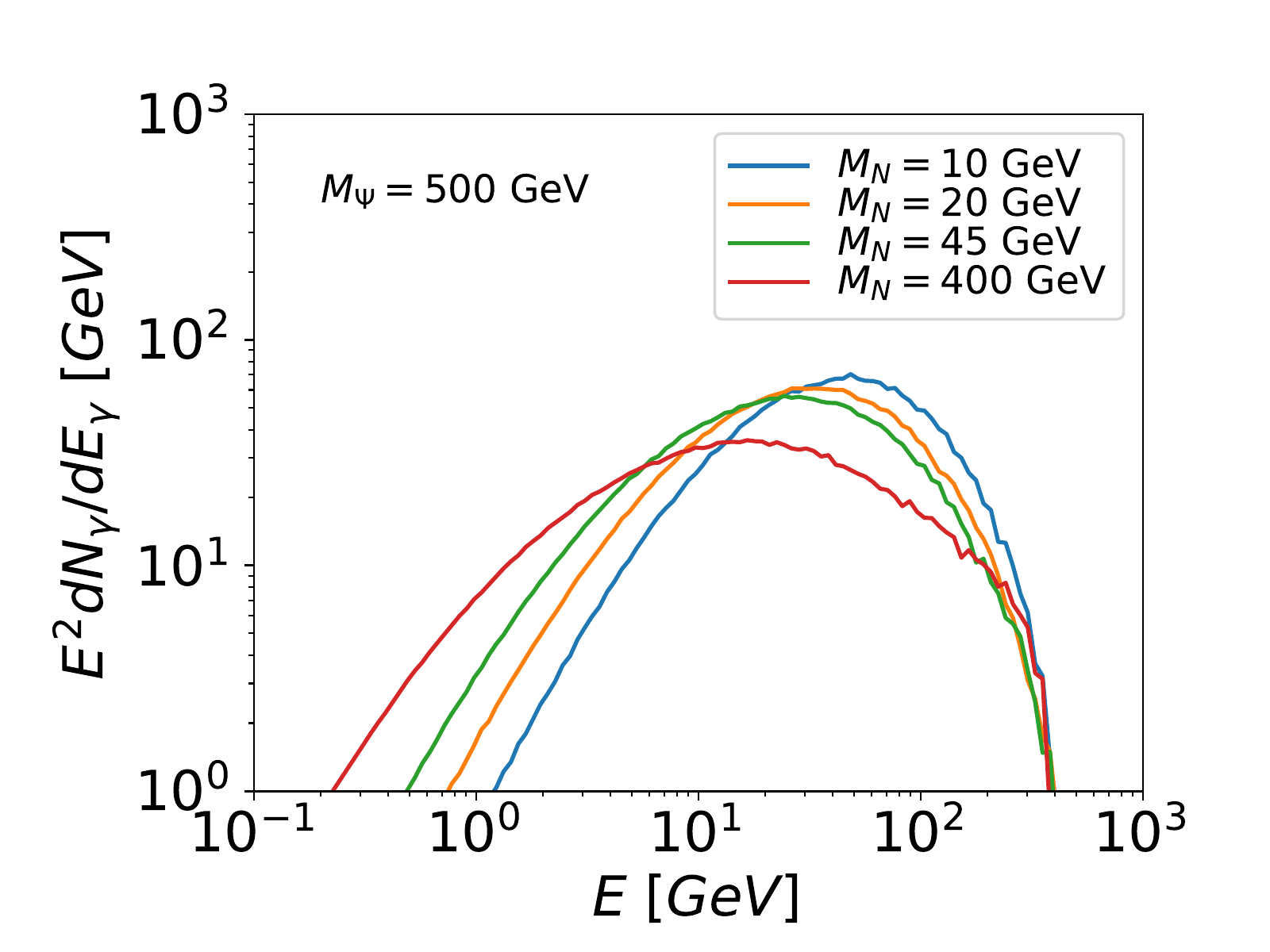}}
\subfigure{\includegraphics[width=76mm]{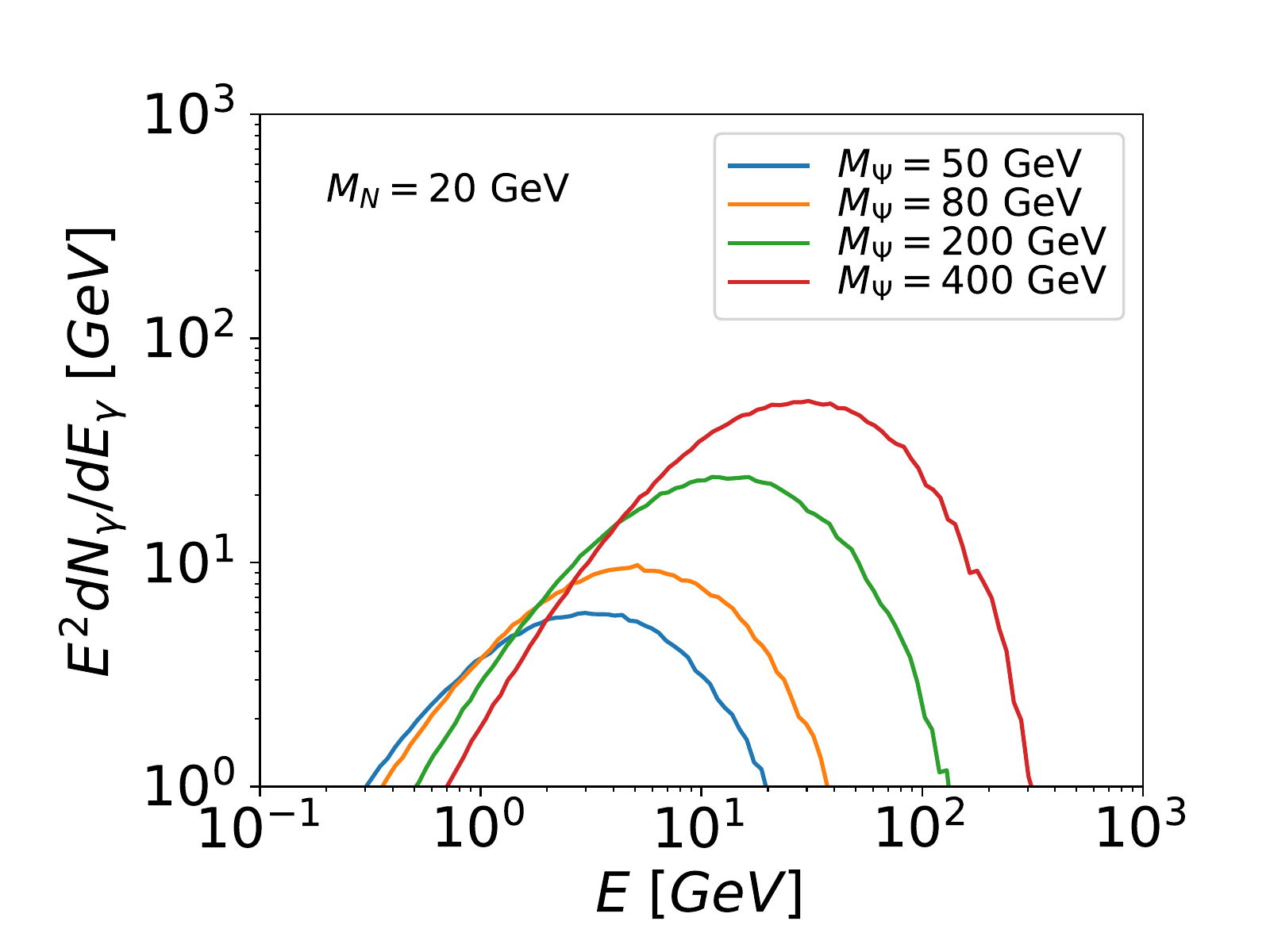}}
\subfigure{\includegraphics[width=76mm]{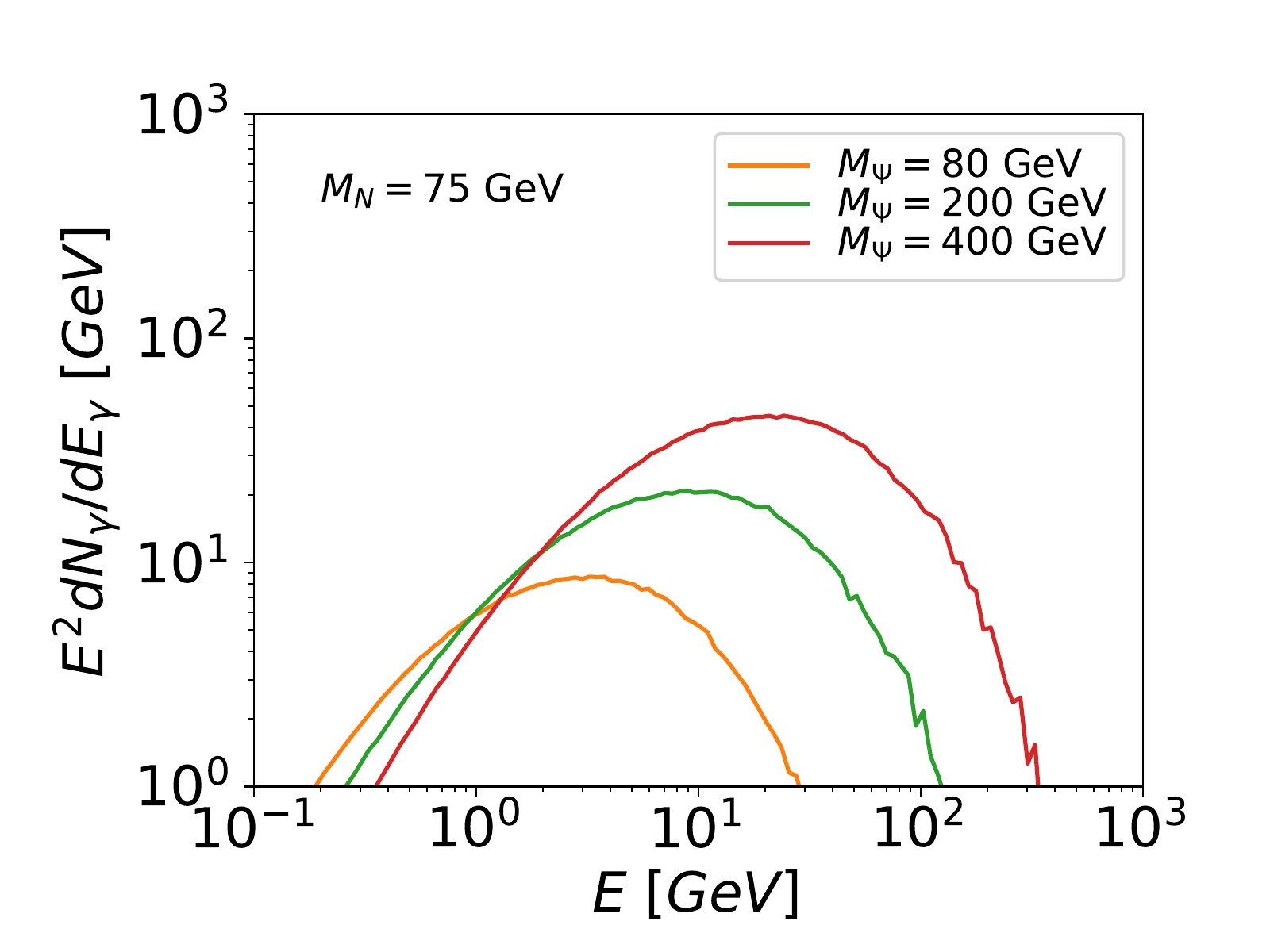}}
\caption{Photon spectrum different DM and sterile neutrino masses. In the upper figures we fix the DM mass and in the lower figures the sterile neutrino one. 
Low DM masses ($\lesssim $ 80 GeV) can fit the Galactic center excess.} 
\label{foton_spectrum}
\end{figure}

To obtain the final state's SM particle spectrum from DM annihilation into sterile neutrinos, $dN/dE$, 
we have used \textit{SPheno v.3.3.8} \cite{Porod:2003um} to determine the decay rates of all the particles, implementing first the model, at the Lagrangian level, using \textit{SARAH v.4.9.1} \cite{Staub:2015kfa, Vicente:2015zba}. Then, we simulate the DM to sterile neutrino annihilation with \textit{MadGraph5 v.2.5} \cite{Alwall:2014hca}, and we use \textit{Pythia v.8.2} \cite{Sjostrand:2007gs} to compute the sterile neutrino decays and its parton shower. 
Our analysis differs from ref.\cite{Batell:2017rol} in that they simulate the decay of sterile neutrino to SM particles in the $N$-rest frame using \textit{SM\_HeavyN\_NLO} model files \cite{Alva:2014gxa,Degrande:2016aje} and boost the final spectrum to the DM rest frame. We have checked that both methods predict similar photon and anti-particle spectra.

In Fig.\ref{foton_spectrum} it is depicted the photon spectrum that we obtain for different DM and $N$ masses: in the upper plots we show the dependence on the sterile neutrino mass for two fixed values of the DM mass, namely $M_{DM} = 50$ GeV, which as we will see in the next section can fit the GCE, and $M_{DM} = 500$ GeV, which do not. We observe that for a given DM mass, the photon spectrum is harder for lighter sterile neutrino. 
 
The reason for this behavior, also observable in the anti-particle spectra, is the boost between the sterile neutrino and DM rest frames, which becomes larger for $M_N \ll M_{DM}$. For instance, an isotropic spectrum with fixed energy $E$ in the sterile neutrino rest frame becomes a box shaped spectrum when boosted to the DM rest frame, of the form \cite{Agrawal:2014oha} 
\begin{equation}
\frac{dN}{dE'} = 
\frac{1}{2\gamma\beta \sqrt{E^2-m^2}} \, 
\theta(E'- E_-) \theta(E_+ - E')
\end{equation}
where $E_\pm = \gamma (E \pm \beta \sqrt{E^2-m^2})$, 
$\theta$ is the Heaviside step function, 
$\gamma$ and $\beta$ are the boost parameters, with 
$\gamma = m_{DM}/m_N$, and $m=0$ for the case of photons
As a consequence, the more boosted the sterile neutrino, the harder the final spectrum. 

In the lower plots the sterile neutrino mass is fixed, and in this case the spectrum is harder for heavier DM mass, as we expected.

In our calculation we have taken only the Yukawa coupling of the sterile neutrino to the first generation of SM leptons non-zero. We have checked that the photon spectrum has little sensitivity to this choice of flavour, in agreement with ref. \cite{Campos:2017odj}; thus the photon spectrum from DM annihilation do not provide insight into disentangling the structure of the sterile neutrino Yukawa couplings.

\begin{figure}[htbp]
\centering
\subfigure{\includegraphics[width=76mm]{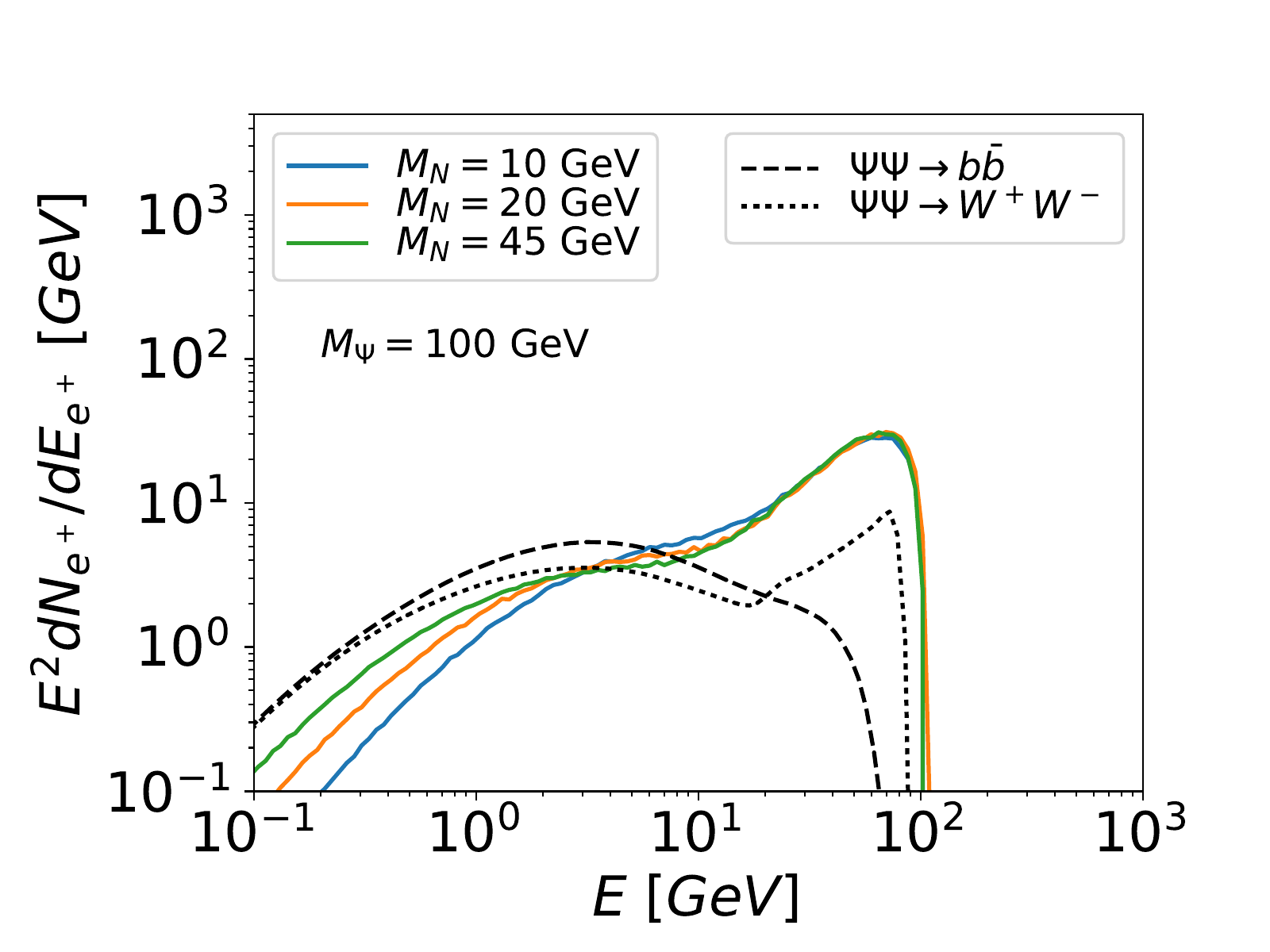}}
\subfigure{\includegraphics[width=76mm]{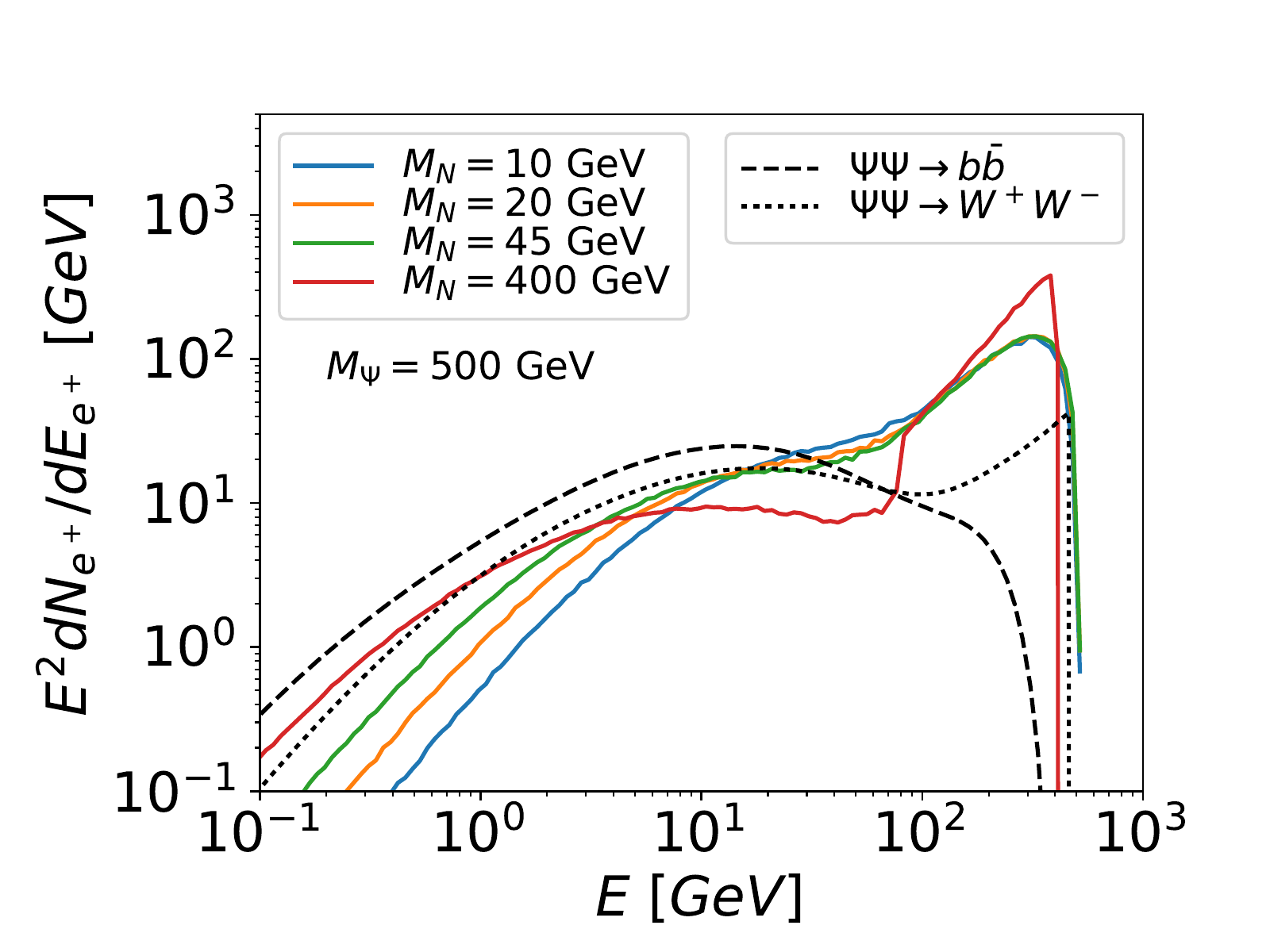}}
\subfigure{\includegraphics[width=76mm]{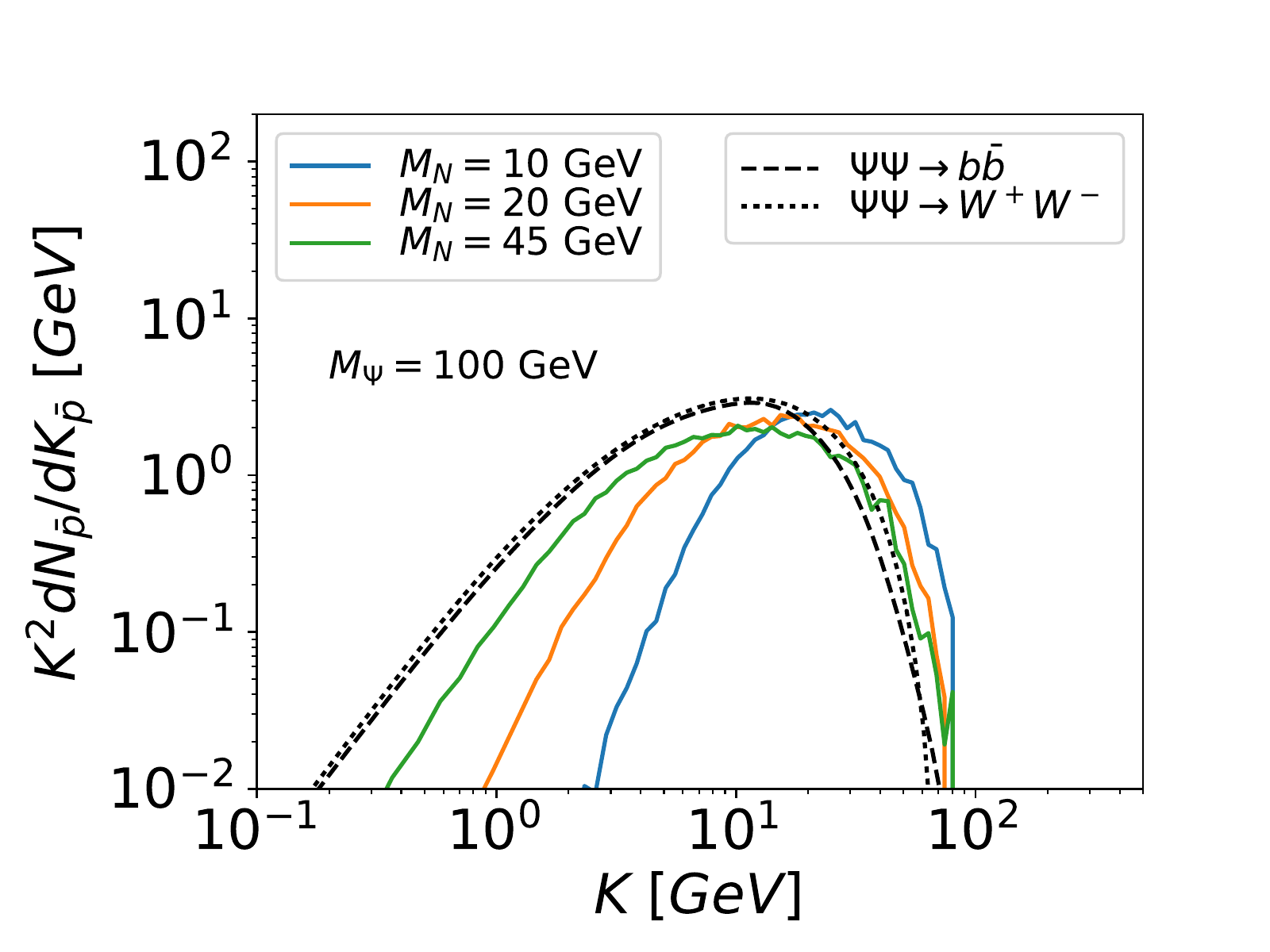}}
\subfigure{\includegraphics[width=76mm]{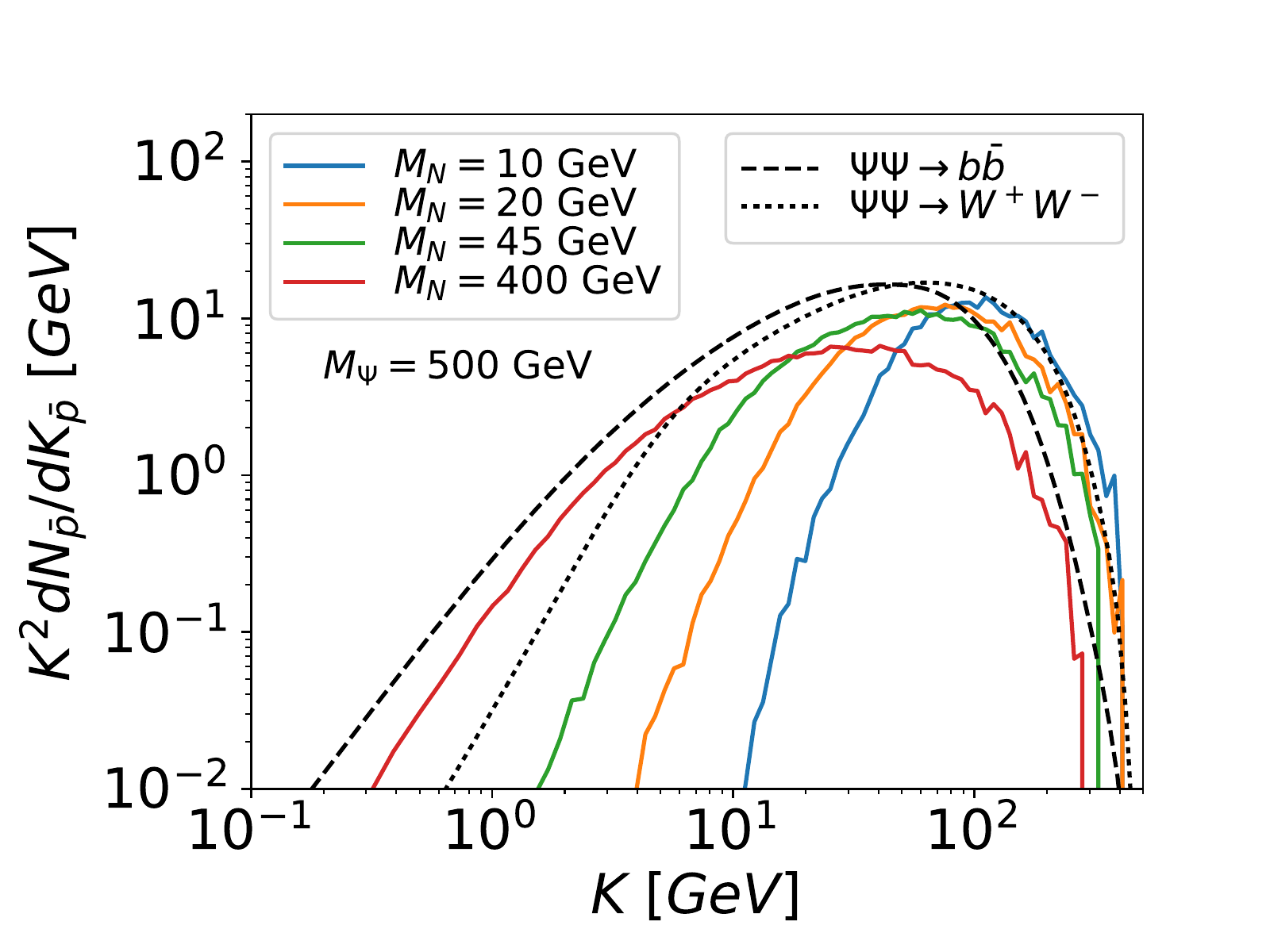}}
\caption{Positron and antiproton spectrum for different DM and sterile neutrino masses compared with a simple case of a DM candidate that annihilates directly to $b\bar{b}$ and $W^+W^-.$} \label{positron_and_antiproton_spectra}
\end{figure}

We have calculated the positron and anti-proton spectra from DM annihilation into sterile neutrinos and its subsequent cascade decay, shown in Fig.~\ref{positron_and_antiproton_spectra}. 
While we find that also the anti-proton spectrum is largely insensitive to the flavour structure of the Yukawa couplings, the peak in the electron spectrum at high energies is only present if the sterile neutrino couples to the $(e,\nu_e)$ doublet, due to a strong component of the reaction $N \rightarrow W e$ which occurs only in this case.

In this work we focus on the $\gamma$-ray probe for several reasons. First of all, we have found that the positron flux generated in the DM sterile neutrino portal can not account for the positron flux observed, for instance, by PAMELA \cite{Adriani:2011xv} and AMS-02 \cite{PhysRevLett.113.221102,PhysRevLett.113.121101}. 
We have used the approximation described in \cite{Cirelli:2010xx} to propagate the positrons and electrons, and obtain the corresponding flux at Earth position. Although the approximation is not very accurate, it is good enough to show that this scenario predicts a positron flux about two orders of magnitude smaller than the measured one, for any value of the DM and $N$ masses; therefore it can not explain the positron excess.

On the other hand, regarding anti-protons, recent analyses of AMS-02 \cite{PhysRevLett.117.091103} data seem to find an excess over the expected background; however a careful study would require a complete fit of both the cosmic ray propagation and DM parameters, which is beyond the scope of this work. 
Nevertheless, as an illustration we plot in Fig.~\ref{positron_and_antiproton_spectra} the anti-proton spectra for several values of the DM and $N$ masses,
together with the spectra corresponding to DM annihilation into $W W$ and $b \bar{b}$ for comparison. 
In Sec.\ref{sec:dSphs}
we will also estimate which part of the parameter space could be excluded by AMS-02 anti-proton data.

Finally, light neutrinos are also produced in DM annihilation, and IceCUBE can set constraints on the cross section to neutrinos, but current limits are about three orders of magnitude above the flux predicted within the sterile neutrino portal scenario \cite{Escudero:2016tzx}.

Note that one can also constrain the sterile neutrino portal using the CMB anisotropy measurements, which are sensitive to DM annihilation during the cosmic dark ages. Specially if the annihilation products contain energetic electrons and photons, when these particles are injected into the plasma will modify the ionization history, leading to observable changes in the temperature and polarization anisotropies. These constraints have been estimated in \cite{Escudero:2016ksa}, and explicitly calculated in 
\cite{Batell:2017rol}, and they exclude DM masses below $\sim$ 20 GeV, irrespective of the value of $M_N$. Therefore, such CMB bounds are weaker than the ones from Fermi-LAT dSphs that we discuss in Sec. \ref{sec:dSphs}.

\section{Analysis of the Galactic Center gamma-ray Excess within the sterile neutrino portal} \label{sec:gc}

The Fermi-LAT has boosted significant advances in our knowledge of the gamma-ray sky over the last few years.
Regarding DM properties, if it is a weakly interacting particle (WIMP) we expect that its annihilation in dense regions of the Universe, such as the our Galactic Center or the DM rich dSphs, will produce a significant flux of SM particles. High energy gamma rays are particularly interesting, since the signal can be traced back to the source, providing information about the location of the DM reaction.
Several studies of the Fermi-LAT data show that the Galactic center is brighter than predicted by conventional models of interstellar diffuse $\gamma$-ray emission \cite{Goodenough:2009gk, Vitale:2009hr, Hooper:2010mq, Gordon:2013vta, Hooper:2011ti, Daylan:2014rsa, 2011PhLB..705..165B, Calore:2014xka, Abazajian:2014fta,Zhou:2014lva,TheFermi-LAT:2015kwa, 1704.03910}, tuned with Galactic plane data and point source catalogs. In a recent analysis by the Fermi-LAT collaboration \cite{TheFermi-LAT:2017vmf}, it has been found that the GCE is a sub-dominant component (10\%) of the observed flux, with a spectral energy distribution peaked at about 3 GeV, slightly shifted towards higher energies than in previous studies. 
We consider the GCE obtained in the so-called Sample Model of ref. \cite{TheFermi-LAT:2017vmf}, and perform the fits using the covariance matrices derived in \cite{Achterberg:2017emt}.

Notice however that the origin of the GCE is still unclear: in addition to the DM explanation,  it could be due to the emission of a population of unresolved point sources \cite{Abazajian:2010zy,Bartels:2015aea,Lee:2015fea,Fermi-LAT:2017yoi,Caron:2017udl}, or cosmic-ray particles injected in the Galactic center region, interacting with the gas or radiation fields \cite{Cholis:2015dea}. In fact, the excess could have different origins below and above $\sim$ 10 GeV \cite{TheFermi-LAT:2017vmf, Linden:2016rcf}: the high energy tail may be due to the extension of the Fermi bubbles observed at higher latitudes, while the lower energy ($<$ 10 GeV) excess might be produced by DM annihilation, unresolved millisecond pulsars, or both.
In conclusion, the interpretation of the GCE as a signal of DM annihilation is not robust, but it can not be ruled out either \cite{TheFermi-LAT:2017vmf, Caron:2017udl}. 

In general, the interpretation of the low energy GCE as originated by DM annihilation is not easy to reconcile with DM direct detection constraints, since in particular models 
the region able to reproduce the excess is already excluded by current experiments: for instance in the context of the minimal supersymmetric standard model, DM can only account for a $\sim$ 40 \% of the low energy ($E < $ 10 GeV) GCE \cite{Achterberg:2017emt}. In our sterile neutrino portal scenario direct detection limits can be easily avoided, provided the mixing angle between the SM Higgs and the dark scalar is small enough; since the relic abundance is determined by the DM annihilation into sterile neutrinos, it is possible to obtain the correct one independently of such mixing. 
In fact, for this reason DM indirect searches are the most promising way to constrain this scenario. See also \cite{Casas:2017rww}, where an extended scalar-singlet Higgs portal model is shown to provide an excellent fit to the GC excess, evading strong direct detection constraints by adding a second (heavier) singlet scalar in the dark sector. 

In our analysis we assume that there are two distinct sources for the GCE: one astrophysical, responsible for the high energy tail of the $\gamma$-ray spectrum, and DM annihilation, that we considerer the only source of the low energy GCE,
\be
\label{3.1}
\Phi = \Phi_{astro} + \Phi_{DM}  \ .
\ee
For the astrophysical component, according to the morphological studies of \cite{TheFermi-LAT:2017vmf} it seems reasonable to consider a continuation to lower Galactic latitudes of the Fermi bubbles.
Given that above 10$^o$ in Galactic latitude the spectral shape of the Fermi bubbles is described by a power low times an exponential cut off \cite{Fermi-LAT:2014sfa}, we assume the 
same form for the astrophysical contribution to the GCE,
\be
\Phi_{astro} = N E^{-\alpha} e^{-E/E_{cut}}
\label{eq:astro}
\ee
We leave $N,\alpha,E_{cut} $ as free parameters in the fit, in order to compare with the values $\alpha = 1.9 \pm 0.2$ and cutoff energy $E_{cut} = 110 \pm 50$ GeV
from the Fermi bubbles, according to the results of ref. \cite{Fermi-LAT:2014sfa}.

For the DM component, the differential flux of photons from a window with size $\Delta\Omega$, is given by \cite{Caron:2015wda}
\begin{equation}
\label{photon_flux}
\frac{d\Phi_{\gamma}}{dE_{\gamma}}(E_{\gamma})=\frac{J}{8\pi M_{DM}^2}\sum_f \langle\sigma v\rangle_f \frac{dN_{\gamma}^f}{dE_{\gamma}}(E_{\gamma}) \ , 
\end{equation}
where the $J$-factor is an astrophysical factor that only depends of the angle of the window size and the DM density profile:
\begin{equation}
\label{eq:J}
J=\int_{\Delta\Omega}{d\Omega}\int{\rho^2_{DM}(s)ds}
\end{equation}
The $J$-factor is an integral of the DM profile over the line of sight. It is very common to adopt the Navarro, Frenk and White (NFW) profile \cite{Navarro:1996gj}. In our case this is the best option because we want to compare our results with the Fermi-LAT data of the GCE and the dSphs, and this is the profile used by the Fermi-LAT Collaboration. The functional form of the NFW profile is:
\begin{equation}
\label{eq:nfw}
\rho_{\Psi}(r)=\rho_s\left(\frac{r}{r_s}\right)^{-\gamma}\left(1+\frac{r}{r_s}\right)^{-3+\gamma} \ ,
\end{equation}
where $r_s = 20$ $kpc$ is the scale radius and $\rho_s$ is the scale density, which is fixed using data at the location of the Sun: at $r_{\astrosun} = 8.5$ $Kpc$, the DM density is $\rho_{\astrosun} = 0.3$ $GeV/cm^3$. We take the central value of $\gamma$ as determined in \cite{TheFermi-LAT:2017vmf}, $\gamma = 1.25 \pm 0.8$.

\begin{figure}[htbp]
\centering
\subfigure{\includegraphics[width=76mm]{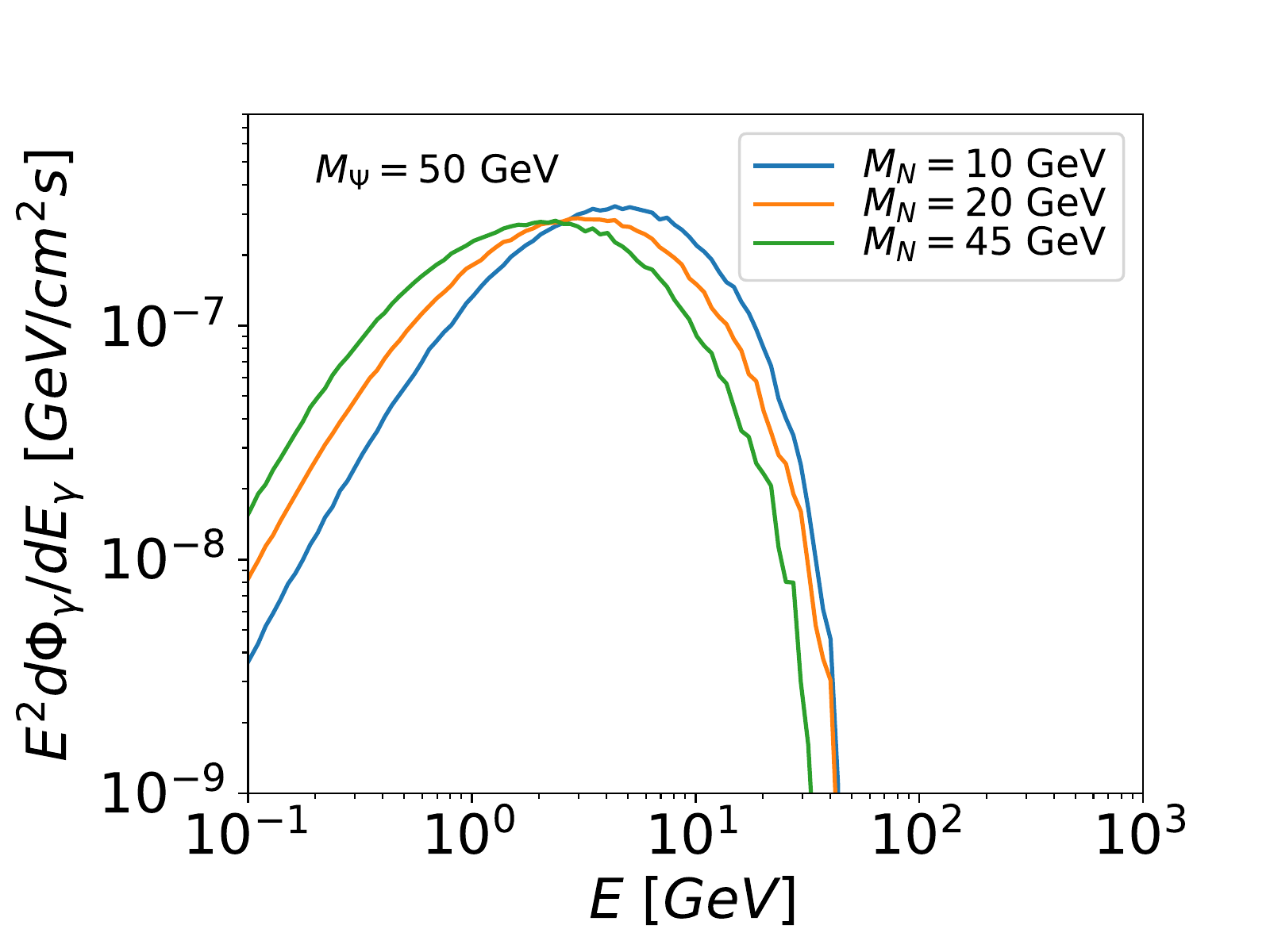}}
\subfigure{\includegraphics[width=76mm]{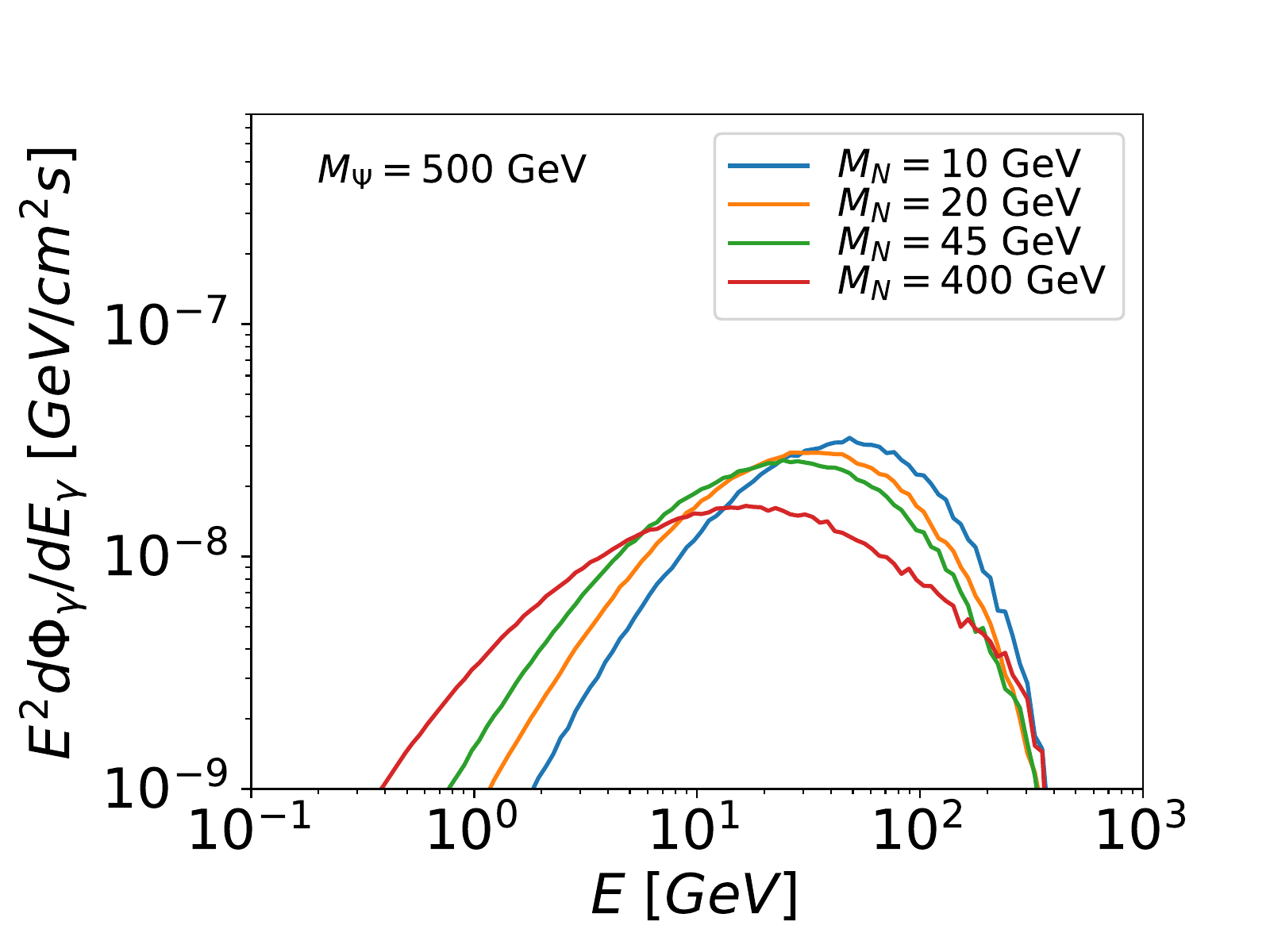}}
\subfigure{\includegraphics[width=76mm]{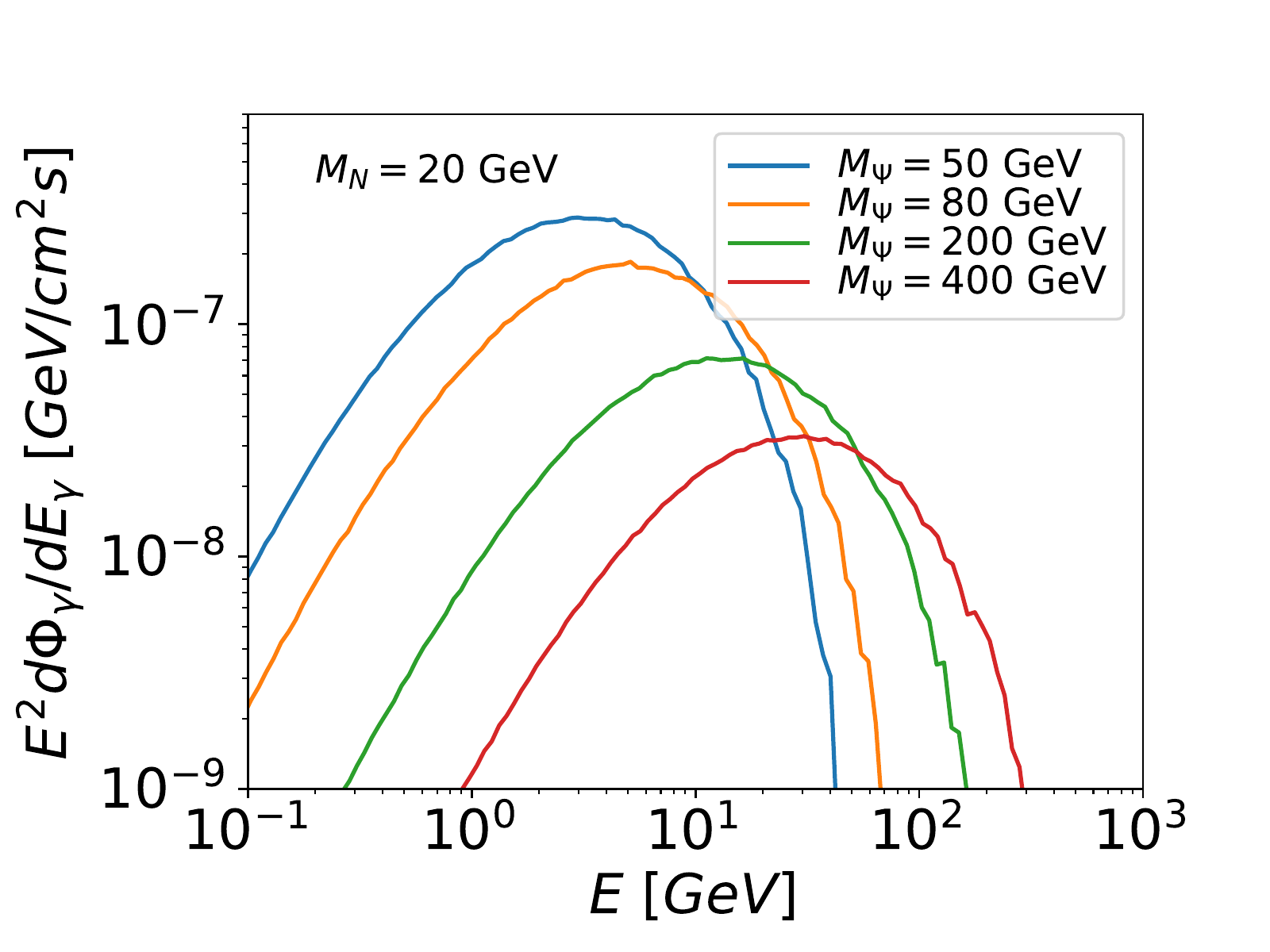}}
\subfigure{\includegraphics[width=76mm]{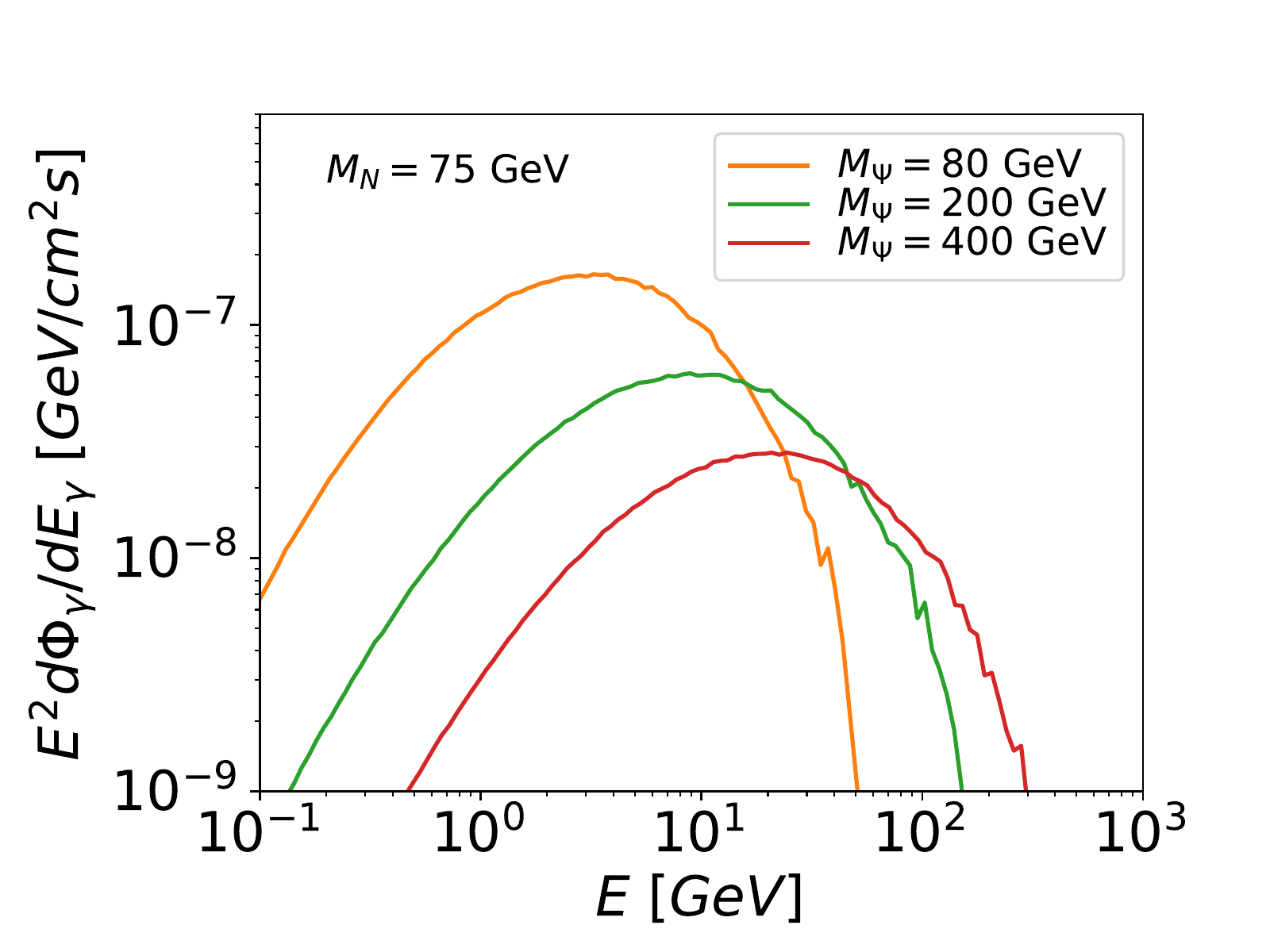}}
\caption{Photon flux for the same points of the parameter space that we chose in figure \ref{foton_spectrum}.} \label{flujo_fotones}
\end{figure}

In Fig.\ref{flujo_fotones} we can see different examples of the photon flux, for the same DM and $N$ masses as in Fig.\ref{foton_spectrum} and thermal annihilation cross section, $\langle \sigma v \rangle = 2.2 \times 10^{-26}$ $cm^3/s$.

\begin{figure}[htbp]
\centering
\includegraphics[width=150mm]{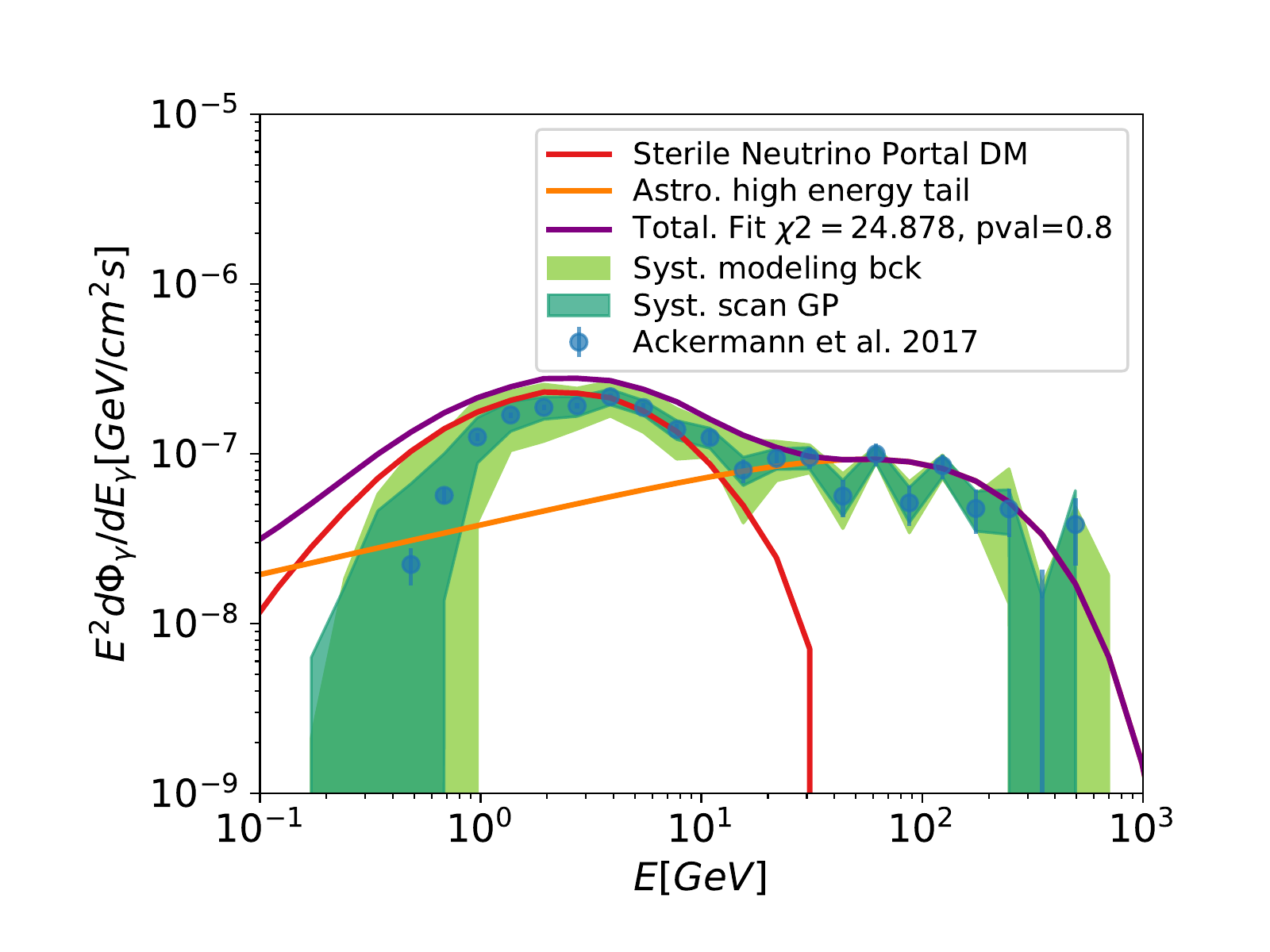}
\caption{Fit to the GCE spectrum (blue dots) by the combination of a power-law with an exponential cutoff, describing the astrophysical sources (orange line), plus the contribution of DM annihilation, as given by dark matter annihilation into sterile neutrinos (red line). The purple line gives the final prediction of the model. The dark green band represents the diagonal of the covariance matrix due to excesses along the Galactic Plane, obtained using the same procedure as for the GCE~\cite{Achterberg:2017emt}. The light green band is the diagonal of the covariance matrix from variations in the GCE due to uncertainties in modelling diffuse emission from ref.~\cite{1704.03910}}.
\label{best_value_GCE}
\end{figure}

We perform a seven parameters fit: $N, \alpha, E_{cut} $ for the astrophysical flux and $J$, $\langle\sigma v\rangle$, $M_\Psi$, $M_N$ for the DM contribution. The quality of the fit is evaluated by constructing the $ \chi^2$ estimator:
\begin{equation}
\label{eq:chi2}
\chi^2=\sum_{i,j} (\Phi_i^{obs} - \Phi_i^m) \Sigma_{i,j}^{-1} (\Phi_j^{obs} - \Phi_j^m)  \ , 
\end{equation}
where $i$ is the energy bin label, $\Phi_i^m$ is the predicted flux for a model, determined by the six free parameters, $\Phi_i^{obs}$ is the flux in the Sample model (light blue points of Fig.\ref{best_value_GCE}) and $\Sigma_{i,j}^{-1}$ is the inverse of the covariance matrix, calculated in \cite{Achterberg:2017emt}. Thus the derived information on the GCE spectrum in \cite{TheFermi-LAT:2017vmf} is contained in $\Phi_i^{obs}, \Sigma_{i,j}^{-1}$.

Notice that since the functions used to fit the GCE are not linear, one can not use the reduced $\chi^2$ to calculate p-values. Instead we perform the following procedure \cite{Achterberg:2017emt}:

1. For each point of the DM model, ($\langle\sigma v\rangle, M_\Psi, M_N$), we vary the astrophysical parameters $N,\alpha,E_{cut}$, as well as the $J$-factor to account for its uncertainties \footnote{We consider one order of magnitude variation in the $J$-factors.},so  as to find the best fit to the data, $\Phi^m_{best}$.

2. We create a set of 100.000 pseudo-random data normal distributed with mean at $\Phi^m_{best}$, according to $\Sigma_{i,j}^{-1}$.

3. We compute $\chi^2$ between $\Phi^m_{best}$ and each of the 100.000 pseudo-random data created in 2.

4. We create a $\chi^2$ distribution using the values from 3.

5. The integrated $\chi^2$ distribution up to the best-fit-$\chi^2$ to the  actual data gives the p-value of the model. 

\begin{figure}[htbp]
\centering
\includegraphics[width=150mm]{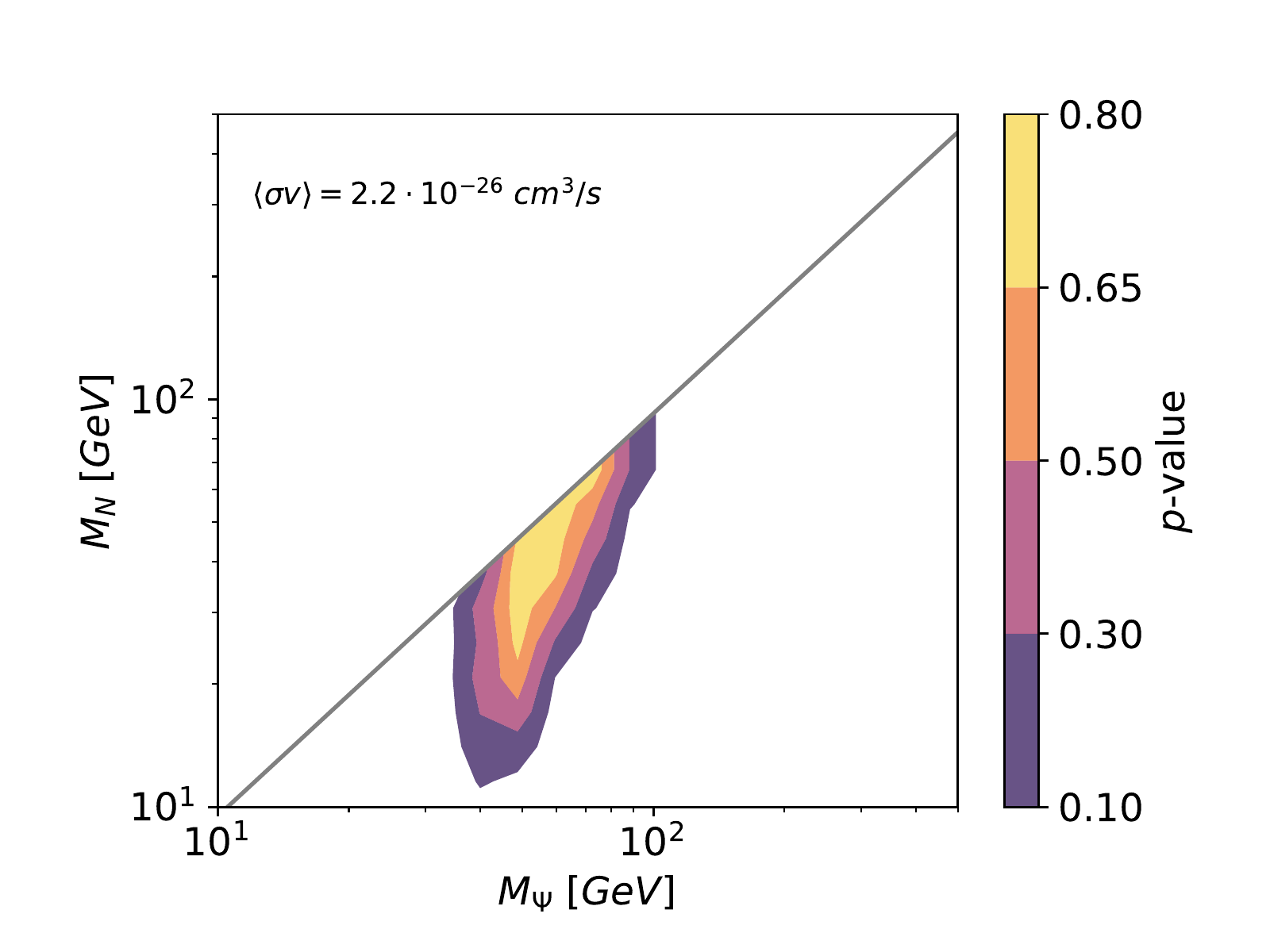}
\caption{The color shows the parameter space region in which the model predicts a $\gamma$-ray flux compatible with the Galactic Center excess for a fixed $\langle \sigma v \rangle = 2.2 \times 10^{-26} {\rm cm^3/s}$. For the fit we use the combined DM and astrophysical components, eq.~(\ref{3.1}).} \label{GCE}
\end{figure}

Due to the uncertainties on the $J$-factors, there is a degeneracy between $J$ and $\langle\sigma v\rangle$, so that a very good fit can be obtained  for 
$\langle\sigma v\rangle$ in the range $0.2 \lesssim
\langle\sigma v\rangle/\langle\sigma v\rangle_{thermal} \lesssim 1.5 $.
As a result, in the best fit point the value of 
$\langle\sigma v\rangle$ is not unambiguously determined, and we have chosen to present the results for 
the thermal one, 
$\langle \sigma v \rangle_{thermal}  = 2.2 \times 10^{-26} {\rm cm^3/s}$  because this is the thermal cross section consistent with the observed DM density.

Fig.~\ref{GCE} shows the different p-values in the ($M_\Psi, M_N$) plane. Notice that it is only possible to fit the GC excess in the low mass region for the DM particle and the sterile neutrinos, more precisely within the range of mass 30-100 GeV for both particles.
From the photon fluxes depicted in Fig.~\ref{flujo_fotones} we can see that increasing the sterile neutrino mass leads to less energetic $\gamma$-rays, while increasing the DM mass produces the contrary effect, the $\gamma$-rays are more energetic. 
On the other hand, the flux decreases for heavier DM, since there are fewer particles contributing. 
These features explain the shape of the fitting regions depicted in Fig.~\ref{GCE}. 

As already mentioned, the fit in Fig.~\ref{GCE} used the new GCE data reported by the Fermi-LAT Collaboration \cite{1704.03910}. The reference \cite{Batell:2017rol} provides an excellent analysis of a previous estimation of the GCE in \cite{Calore:2014xka}. We find a larger parameter space allowed to fit the GCE data than in \cite{Batell:2017rol} mainly because of the broader systematic uncertainties in the GCE estimation that we used and our inclusion of an extra astrophysical component to model the GCE.

Fig.~\ref{best_value_GCE} shows the photon flux for our best fit point of the parameter space. Combining the astrophysical and the DM component, as given in eq.~(\ref{3.1}), we obtain that the best fit point is $(M_{\Psi}, M_N)=(55.1,51.4)$ GeV for the DM component. For this point the best values of the astrophysical parameters are $(N,\alpha,E_{cut})=(3.81 \times 10^{-8} \, {\rm GeV^\alpha},1.7,187.8 \, {\rm GeV})$. 
We obtain a very good fit, $\chi^2 = 24.9$ for 27 energy bins, which corresponds to a p-value = 0.78.

\section{Constraints from indirect detection: gamma rays from dSphs and anti-proton data}
\label{sec:dSphs}

In the previous section, we have analyzed the photon flux from DM annihilation into sterile neutrinos, and its impact in the GCE. In this section, we will constrain the parameter space with the non-detection of dSphs by the Fermi LAT. Given the large diversity of photon spectra in the DM sterile neutrino portal to DM scenario, see fig. \ref{foton_spectrum}, we can not use the limits presented in the Fermi-LAT Collaboration publications, as they are for some particular annihilation channels \cite{Fermi-LAT:2016uux}. Therefore, we use \textit{gamLike v.1.0} \cite{Workgroup:2017lvb}, a software that evaluates the likelihoods for $\gamma$-ray searches using the combined analysis of 15 dSphs from 6 years of Fermi-LAT data, processed with the Pass-8 event-level analysis. \textit{gamLike} calculates the Poisson likelihood following the method described in \cite{Ackermann:2015zua}. First of all we define the J-factor likelihood:
\begin{equation}
\label{Jfactor_likelihood}
\mathcal{L}_J(J_i|J_{obs,i,},\sigma_i)=\frac{e^{-(\log_{10}(J_i) - \log_{10}(J_{obs,i}))^2/2\sigma_i^2}}{\ln(10)J_{obs,i}\sqrt{2\pi}\sigma_i} \,
\end{equation}
where $J_{obs,i}$ is the measured $J$-factor with error $\sigma_i$ in each dSphs $i$ and $J_i$ is the true $J$-factor value. We then define the combined likelihood of all dSphs in the form:
\begin{equation}
\label{Jfactor_likelihood_2}
\mathcal{L}_i(\mu,\theta_i|D_i)=\prod_j \mathcal{L}_i(\mu,\theta_i|\mathcal{D}_{i,j}) \,
\end{equation}
where $\mu$ are the parameters of the DM model, $\theta_i$ accounts for the set of nuisance parameters from the LAT study and $J$-factors of the dSphs, and $D_i$ is the $\gamma$-ray data set. 

Using these ingredients we perform a test statistic (TS) to obtain 90\% C.L. upper limits on the DM annihilation cross section. Such bounds are derived by finding a change in the log-likelihood:
\begin{equation}
\label{loglikehood}
TS= -2 \ln\frac{\mathcal{L}(\mu_0,\hat{\theta}|\mathcal{D})}{\mathcal{L}(\hat{\mu},\hat{\theta}|\mathcal{D})} \,
\end{equation}
where $\mu_0$ are the parameters of the no DM case (when we do not have $\gamma$-rays in our model) while $\hat{\mu}$ and $\hat{\theta}$ are the parameters for the point we want to analyze.

If $TS > 2.71$ the parameter space point is excluded because it is not compatible with the background at 90\% C.L. Using this method we can find the exclusion line in the plane $M_{\Psi} - M_N$.

\begin{figure}[htbp]
\centering
\includegraphics[width=150mm]{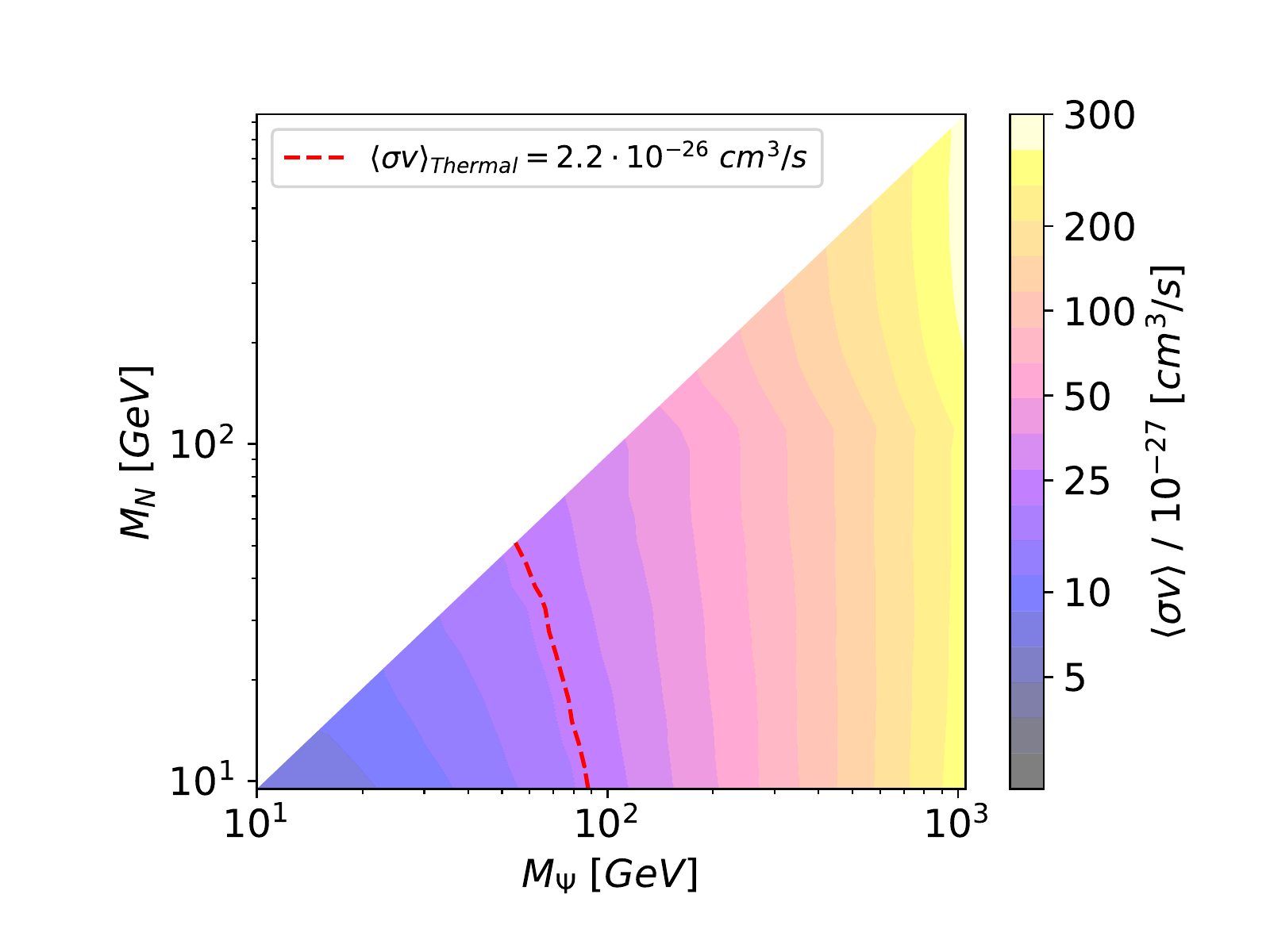}
\caption{dSphs exclusion limit defined using the compatibility with the background. In this plot the color code shows the constrain for different values of $\langle\sigma v\rangle$.} \label{Dwarf_limits}
\end{figure}

We can see in Fig.~\ref{Dwarf_limits} the contour limits, corresponding to different $\langle\sigma v\rangle$ values. The region to the left of the corresponding curve is excluded at 90\% C.L. 
We show as a red-dashed line the limit for a thermal annihilation cross-section, which in principle is the one needed to obtain the observed DM relic abundance within the sterile neutrino portal scenario under study. We find that DM masses $M_\Psi < 60$ GeV are excluded, in agreement with \cite{Batell:2017rol}, a somehow weaker limit than the one obtained in \cite{Campos:2017odj}.

Note however than in some cases the dark sector (including the sterile neutrino) could be at a different temperature than the SM, so that a larger freeze-out annihilation cross section is required to fit the observed DM abundance \cite{Tang:2016sib,Bernal:2017zvx}. Therefore a larger region of the parameter space ($M_\Psi, M_N$) is excluded in such cases.

\begin{figure}[htbp]
\centering
\includegraphics[width=150mm]{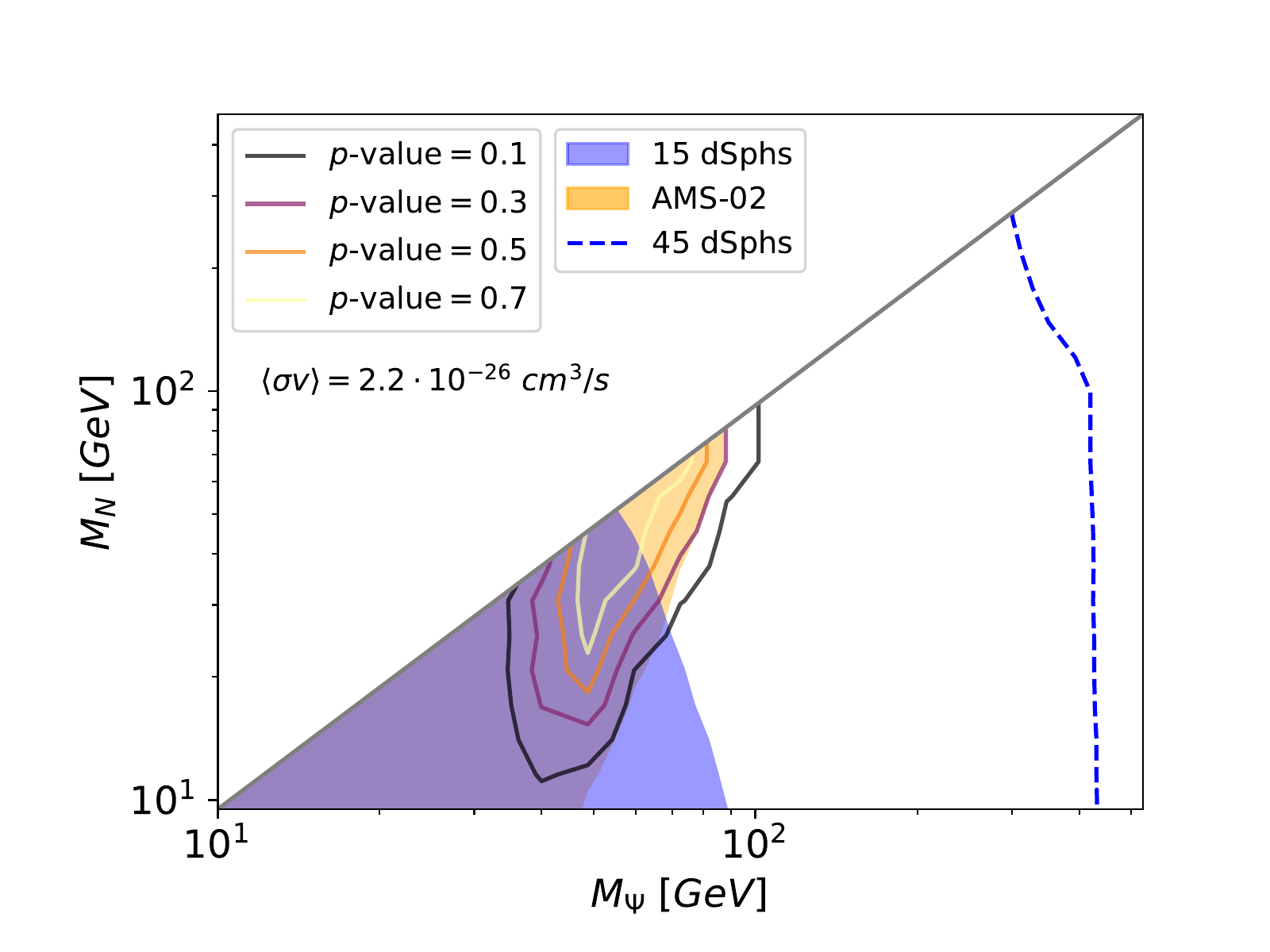}
\caption{Region of the parameter space that fit the GCE combined with the dSphs and AMS-02 anti-protons constrain for a thermal value of the $\langle\sigma v\rangle$.} 
\label{dwarf}
\end{figure}

Focusing on the standard thermal annihilation cross-section, we next analyze the impact of the dSphs constraints on our fit of the GCE. In Fig.~\ref{dwarf} we plot both results, and we can see that the dSphs limit disfavours the low DM mass region of our fit of the GCE, although a sizable range of ($M_\Psi, M_N$) able to fit the GCE, remains allowed.

We expect that the sensitivity of the Fermi-LAT telescope will improve significantly in the next years by, among other reasons, the potential discoveries of new ultra-faint dwarf galaxies \cite{Charles:2016pgz}. Using a similar analysis to \cite{Horiuchi:2016tqw}, we estimate that in 15 years of data taking Fermi-LAT will have 3 times more dSphs discovered (45 dSphs) and considering that the point spread function (PSF) sensitivity for the Fermi-LAT instrument increases approximately as the square-root of the observation time (this is a conservative estimate), the Fermi-LAT constraints will improve by a factor of $(\sqrt{15}/\sqrt{6})\times 3 \simeq 5$. In Fig.~\ref{dwarf} we show the impact of this prospect (dashed blue line): the region to the left of this line  will be potentially excluded in the next years by Fermi-LAT, including the GCE fit area (if we assume that all low energy GCE is due to DM annihilation).

Finally, we roughly estimate the effect of anti-proton 
data from AMS-02 on the sterile neutrino portal allowed parameter space. The derivation of these bounds suffer from large uncertainties, one of them being 
that the propagation parameters in the traditional MIN-MED-MAX schemes are determined by old Cosmic Ray data, and they are not necessarily guaranteed to describe the current status; indeed for instance the MIN propagation scheme is seriously disfavored \cite{Giesen:2015ufa} by the 
preliminary anti-proton to proton ratio reported by AMS-02.
However, the MED scheme seems to provide a reasonable fit to the data, at least in the low energy region, so we have considered it to assess the region that could be excluded by AMS-02 data.
Therefore our results should be taken as an indication of 
the parameter space that would be excluded by a complete fit of the cosmic ray propagation  and DM parameters. We do not attempt here to explain the excess at high anti-proton energies.

We estimate the total flux of anti-protons within our model as the sum of the best fit of the background in the MED scheme \cite{Giesen:2015ufa}, $\Phi_{\bar{p},bkg}(K)$,
plus the DM contribution, i.e., 
$\Phi_{\bar{p}}(K_i,M_{\Psi},M_N) = \Phi_{\bar{p},bkg}(K_i) + \Phi_{\bar{p},\Psi}(K_i,M_{\Psi},M_N)$. 
Then, we calculate the ratio between this flux and the 
proton flux data $\Phi_{p}(K_i)$ from AMS-02 \cite{PhysRevLett.114.171103}, 
in order to compare it with the last experimental data on 
the  
anti-proton-to-proton flux ratio $R(K_i) \pm \sigma_i$, also obtained by the AMS-02 experiment
\cite{PhysRevLett.117.091103}.

\begin{figure}[htbp]
\centering
\subfigure{\includegraphics[width=76mm]{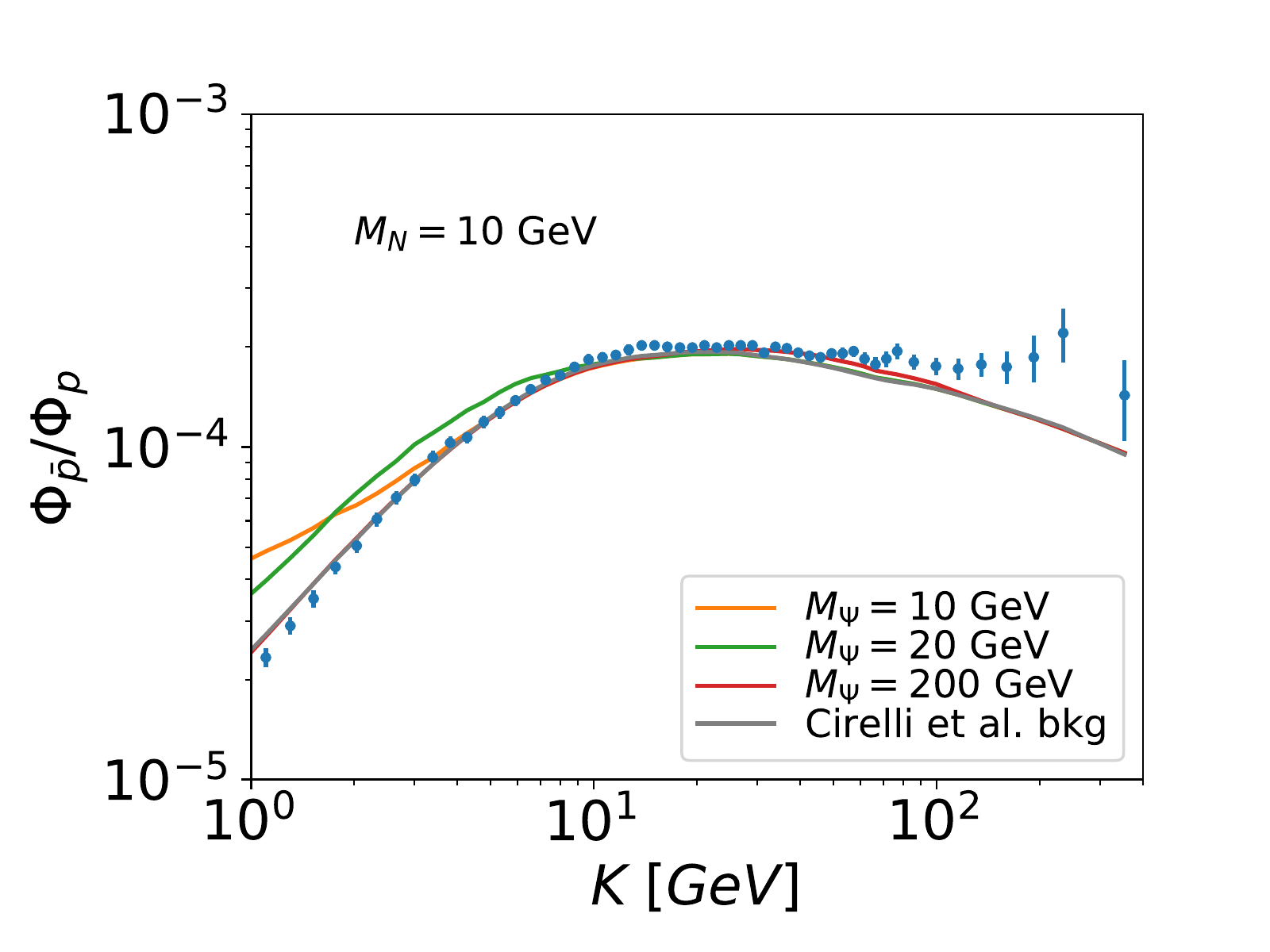}}
\subfigure{\includegraphics[width=76mm]{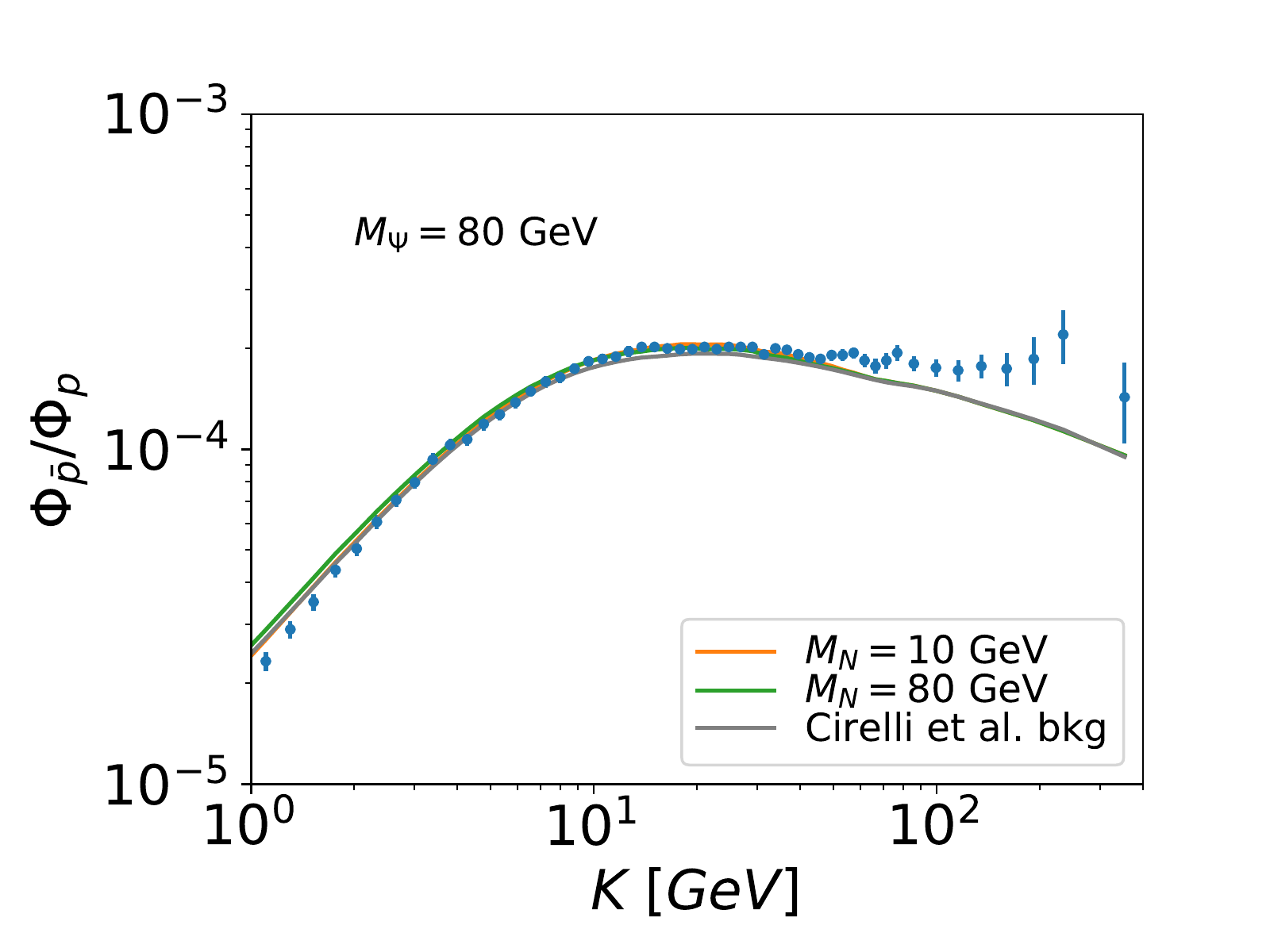}}
\caption{Total anti-proton-to-proton flux ratio for different points of the parameter space compared with the background contribution in the MED propagation scheme (gray line in both plots). Blue dots correspond to AMS-02 data \cite{PhysRevLett.117.091103}. In the left panel we can see the effect of the variation of $M_\Psi$, whereas in the right one the effect of the variation of $M_N$}.
\label{antiprotons}
\end{figure}

In Fig.~\ref{antiprotons} we show the anti-proton-to-proton flux ratio from the 
background (gray line) and for different $(M_\Psi,M_N)$ points as calculated in the MED propagation scheme, together with the recent AMS-02 data. 
Notice that since the data is in agreement or below  the astrophysical background model at low values of the anti-proton kinetic energy $K$, points of the parameter space leading to larger ratios are disfavored. 

Now, for each point of our parameter space 
$(\langle \sigma v \rangle, M_\Psi, M_N)$
we construct the estimator:
\begin{equation}
\chi^2 = \sum_i \left[\frac{R(K_i)-\Phi_{\bar{p}}(K_i,M_{\Psi},M_N))/\Phi_{p}(K_i)}{\sigma_i}\right]^2
\ ,
\end{equation}
where $i$ denotes the energy bins, and $\sigma_i$ the corresponding uncertainty on the flux ratio. 
Denoting $\chi_0^2$ the minimum chi-squared of the background-only case from \cite{Giesen:2015ufa}, we can define  the limit 
on $\langle \sigma v \rangle$ for each point ($M_{\Psi},M_N$) using the condition:
\begin{equation}
\chi^2(\langle \sigma v \rangle, M_{\Psi},M_N) - \chi_0^2 \leq 4
\end{equation}
Note that in this derivation we have used the Einasto 
DM density profile, since it is the one employed by AMS-02.

In Fig.~\ref{dwarf} we show the impact of the anti-proton AMS-02 data using the MED propagation scheme on the 
sterile neutrino portal parameter space.
The orange region corresponds to the ($M_{\Psi},M_N$) points for which the limit on 
$\langle \sigma v \rangle$ obtained in the way described above is 
$\leq 2.2 \times 10^{-26} {\rm cm^3/s}$.
Our results for the MED propagation scheme agree with ref.\cite{Batell:2017rol}, 
where a similar analysis has been performed. As noticed there, the constrains from anti-proton are complementary to the dSphs ones, and for a fixed $M_\Psi$ they disfavour the high $M_N$ region of the GCE fit, since heavier sterile neutrinos produce a larger anti-proton flux at low kinetic energies $K$. However the astrophysical uncertainties are still very large, as has been shown in \cite{Batell:2017rol} by using different propagation schemes and DM density profiles, as well as varying the $J$-factors.

\section{Conclusions}\label{sec:con}

The DM relic abundance could be determined by the freeze-out of DM interactions with sterile neutrinos, which in turn generate light neutrino masses via the seesaw type I mechanism; this is the so-called sterile neutrino portal to DM.
Generically such scenario is challenging to test at colliders and easily evades DM direct searches. However, it can be probed in DM indirect detection experiments, since the sterile neutrinos copiously produced in DM annihilations will subsequently cascade decay into SM final states due to its mixing with the active neutrinos (unless the annihilation cross section is p-wave and therefore it is velocity suppressed at present).

In this work, we focus on the impact of the new analysis of the Fermi-LAT Collaboration of the Galactic Center region, based on the reprocessed Pass 8 event data, which confirms the existence of a $\gamma$-ray excess peaked at $\sim$ 3 GeV. We assume that annihilation of DM into sterile neutrinos is the main contributor to the low energy photon flux of the GCE (photon energy $<$ 10 GeV). The high energy tail of the GCE ($>$ 10 GeV) could be due to an astrophysical component, which we model as a power law with an exponential cut-off, eq.~(\ref{eq:astro}). Although the interpretation of the GCE as DM annihilation is still under debate, it is worth to explore whether a complete particle physics model can account for it.
  
We perform a model-independent analysis within the sterile neutrino portal scenario. Indeed, our results only depend on the thermally averaged DM annihilation cross section into sterile neutrinos, which we fix to $\langle \sigma v \rangle = 2. 2 \times 10^{-26} {\rm cm}^3/s$, and the DM and sterile neutrino masses, $(M_\Psi, M_N)$. Therefore, our analysis can be extended to any model able to reproduce the thermal DM annihilation cross section into sterile neutrinos.

We find that the sterile neutrino portal to DM provides an excellent fit to the GCE: $\chi^2 = 24.9$ for 27 energy bins (p-value = 0.78). The best fit corresponds to $(M_\Psi, M_N) = (55.1,51.4)$ GeV.

We then check the compatibility of these results with the limits from Fermi-LAT Pass 8 data on the dSphs positions
and anti-proton data from AMS-02. Fig.~\ref{dwarf} summarizes our main findings. We see
that there is a sizeable region in the $(M_\Psi, M_N)$ plane able to contribute significantly to the GCE and allowed by the dSphs constraints. Indeed, the dSphs set an stringent limit which excludes DM masses below $\sim$ 50 GeV (90 GeV), for sterile neutrino masses $M_N \lesssim M_{DM}$ ($M_N \ll M_{DM}$). In particular, the above best-fit point to the GCE is allowed. It is worth noticing that shortly further constraints from a larger number of dSphs may be in tension with the explanation of the GCE, under the assumption that a large fraction of the low energy sector of the GCE (below $\approx$ 10 GeV) is due to DM annihilation.

On the other hand, using the MED propagation scheme 
we find that current anti-proton data from AMS-02 already disfavours a large fraction of the $(M_\Psi, M_N)$  region able to account for the GCE; however our analysis is not conclusive,
given the large uncertainties in the anti-proton background estimate and propagation parameters.

 \section*{Acknowledgements} 
 We thank Brian Batell for correspondence about the 
 photon and anti-proton spectrum of sterile neutrino decays.  This work has been partially supported by the European Union projects H2020-MSCA-RISE-2015-690575-InvisiblesPlus and H2020-MSCA-ITN-2015/674896-ELUSIVES,
by the Spanish MINECO under grants FPA2014-57816-P and  SEV-2014-0398,
and by Generalitat Valenciana grant PROMETEOII\-2014/050. The work of GAGV was supported by Programa FONDECYT Postdoctorado under grant 3160153.

\bibliography{DM} 
 \end{document}